\documentstyle[preprint,aps]{revtex}	
\begin{document}
%
%
\preprint{$
\begin{array}{l}
\mbox{UB-HET-95-03}\\[-3mm]
\mbox{UCD--95--21}\\[-3mm]
\mbox{July~1995} \\   [10.mm]
\end{array}
$}
\title{QCD Corrections and Non-standard Three Vector Boson Couplings in
$W^+W^-$ Production at Hadron Colliders}
\author{U.~Baur}
\address{
Department of Physics, SUNY at Buffalo, Buffalo, NY 14260, USA}
\author{T.~Han and J.~Ohnemus}
\address{
Department of Physics, University of California, Davis, CA 95616, USA}
\maketitle
\begin{abstract}
\baselineskip15.pt  
The process $p\,p\hskip-7pt\hbox{$^{^{(\!-\!)}}$} \rightarrow
W^{+} W^{-} + X \rightarrow \ell^+_1 \nu_1 \ell^-_2 \bar \nu_2 + X$
is calculated to ${\cal O}(\alpha_s)$ for general $C$ and $P$ conserving
$WWV$ couplings ($V=\gamma,\, Z$). The prospects for probing
the $WWV$ couplings in this reaction are explored.
The impact of ${\cal O}(\alpha_s)$ QCD corrections and various
background processes on the observability of non-standard $WWV$
couplings in $W^+ W^-$ production at the Tevatron and the Large Hadron
Collider (LHC) is discussed in detail.
Sensitivity limits for anomalous $WWV$ couplings are derived at
next-to-leading order for the Tevatron and
LHC center of mass energies, and are compared to the bounds
which can be achieved in other processes. Unless a jet veto or a cut on
the total transverse momentum of the hadrons in the event is imposed,
the ${\cal O}(\alpha_s)$ QCD corrections and the background from top
quark production decrease the sensitivity of
$p\,p\hskip-7pt\hbox{$^{^{(\!-\!)}}$} \rightarrow
W^{+} W^{-} + X \rightarrow \ell^+_1 \nu_1 \ell^-_2 \bar \nu_2 + X$
to anomalous $WWV$ couplings by a factor two to five.
\end{abstract}
\vskip .2in
\newpage
%
%
\baselineskip 19.pt
\narrowtext

\section{INTRODUCTION}

The electroweak Standard Model (SM) based on an
$\hbox{\rm SU(2)} \bigotimes \hbox{\rm U(1)}$ gauge theory
has been remarkably successful in describing contemporary high
energy physics experiments, however, the three vector boson couplings
predicted by this non-abelian gauge theory remain largely untested. A
precise measurement of these couplings will soon be possible in $W$ pair
production at LEP~II~\cite{LAGRANGIAN,Ber}. With the large data samples
collected in the present Tevatron collider run, and plans for further
upgrades in luminosity~\cite{Brenna},
the production of $W^+ W^-$ pairs at hadron colliders provides an
alternative and increasingly attractive opportunity to study the
$WW\gamma$ and $WWZ$
vertices~\cite{FIRSTWW,BS,WWAC,HWZ,EHLQ}. Recently, the CDF and D\O\
Collaborations reported first measurements of the $WWV$ couplings
($V=\gamma, Z$)
in $W^+W^-$ production at the Tevatron from the data collected
in the 1992 -- 93 run. CDF used the
reaction $p\bar p\to W^+W^-\to\ell^\pm\nu jj$, $\ell=e,\,
\mu$~\cite{CDFWW} to derive limits on anomalous three vector
boson couplings, whereas D\O\ analyzed the dilepton
channels, $p\bar p\to W^+W^-\to\ell_1^+\nu_1\ell_2^-\bar\nu_2$,
$\ell_{1,2}=e,\,\mu$~\cite{DOWW}. In the SM, the $WWV$ vertices are
completely fixed by the $\hbox{\rm SU(2)} \bigotimes \hbox{\rm U(1)}$
gauge structure of the electroweak sector, thus a measurement of these
vertices provides a stringent test of the SM.

In contrast to low energy data and high precision measurements at the $Z$
peak, collider experiments offer the possibility of a direct, and
essentially model-independent, determination of the three vector boson
vertices. Previous theoretical studies on probing the $WWV$ vertices
via hadronic $W^+ W^-$ production have been based on leading-order (LO)
calculations~\cite{FIRSTWW,BS,WWAC,HWZ,EHLQ}.
The prospects for extracting
information on anomalous $WWV$ couplings from decay modes where one of the
$W$ bosons decays into leptons and the second into hadrons,
$W^+W^-\to\ell^\pm\nu jj$, have been discussed in Ref.~\cite{HWZ}. A
detailed discussion of the purely leptonic channels,
$W^+W^-\to\ell_1^+\nu_1\ell_2^-\bar\nu_2$, has not yet appeared in the
literature. In general, the inclusion of anomalous couplings at
the $WW\gamma$ and $WWZ$ vertices yields
enhancements in the $W^+ W^-$ cross section, especially at large values of
the $W$ boson transverse momentum, $p_T^{}(W)$,
and at large values of the $W^+ W^-$ invariant mass, $M_{WW}$.
Next-to-leading-order (NLO) calculations of hadronic $W^+ W^-$
production have shown that the ${\cal O}(\alpha_s)$ corrections
are large in precisely these same regions~\cite{WW,WWFRIX}.
It is thus vital to include the ${\cal O}(\alpha_s)$ corrections when using
hadronic $W^+ W^-$ production to probe the $WW\gamma$ and $WWZ$ vertices.

In this paper, we calculate hadronic $W^+ W^-$ production to ${\cal
O}(\alpha_s)$, including the most general, $C$ and $P$ conserving, anomalous
$WW\gamma$ and $WWZ$ couplings, and discuss in detail the purely leptonic
decay modes, $W^+W^-\to\ell_1^+\nu_1\ell_2^-\bar\nu_2$. Decay channels
where one or both of the $W$ bosons decay into hadrons are not
considered here.
Presently, experiments only place an upper limit on the cross section
for $W^+W^-$ production in hadronic collisions~\cite{CDFWW,DOWW}. With
CDF and D\O\ rapidly approaching their goal of an integrated luminosity
of 100~pb$^{-1}$ in the current Tevatron run, this situation is expected
to change soon~\cite{Doug}.
In the Main Injector Era, integrated luminosities of order
1~fb$^{-1}$ are envisioned~\cite{Brenna,GPJ}, and a sufficient number of
events should be available to commence a detailed investigation of the
$WWV$ vertices in the $W^+W^-\to\ell_1^+\nu_1\ell_2^-\bar\nu_2$ channel,
provided that the background can be controlled. The
prospects for a precise measurement of the $WWV$ couplings in this
channel would further improve if a significant upgrade in luminosity
beyond the goal of the Main Injector could be realized.
With recent advances in
accelerator technology~\cite{GPJ}, Tevatron collider luminosities of
order $10^{33}~{\rm cm}^{-2}~{\rm s}^{-1}$ may become reality
within the next few years, resulting in integrated luminosities of up to
10~fb$^{-1}$ per year (a luminosity upgraded Tevatron
will henceforth be denoted by TeV*). At the CERN Large Hadron
Collider [(LHC), $pp$ collisions  at $\sqrt{s}=14$~TeV~\cite{LHC}], the
$t\bar t$ background needs to be reduced by at least one order of
magnitude in order to utilize the potential of the process $pp\to W^+W^-+X$ to
constrain anomalous gauge boson couplings.

Compared to other processes which are sensitive to the structure of the
$WWV$ vertices,
$W^+W^-$ production has an important advantage. Terms proportional to
the anomalous coupling $\Delta\kappa_V$ in the amplitude (see
Eq.~(\ref{EQ:LAGRANGE}) for a definition of the anomalous couplings)
grow like $\hat s/M_W^2$~\cite{LAGRANGIAN}, where $\hat s$ is the
parton center of mass
energy squared, whereas these terms increase only like $\sqrt{\hat s}/M_W$
in $W^\pm\gamma$ and $W^\pm Z$ production. One therefore expects that
$W^+W^-$ production is considerably more sensitive to $\Delta\kappa_V$ than
$p\,p\hskip-7pt\hbox{$^{^{(\!-\!)}}$} \to W^\pm\gamma,\,W^\pm Z$.

To perform our calculation, we use the Monte Carlo method for NLO
calculations described in Ref.~\cite{NLOMC}. The leptonic decays
of the $W$ bosons are included using the narrow width approximation.
With the Monte Carlo method, it
is easy to calculate a variety of observables simultaneously and to
implement experimental acceptance cuts in the calculation. It is also
possible to compute the ${\cal O}(\alpha_s)$ QCD corrections
for exclusive channels,
{\it e.g.}, $p\,p\hskip-7pt\hbox{$^{^{(\!-\!)}}$} \rightarrow W^+ W^-
+0$~jet. Apart from anomalous contributions to the $WW\gamma$ and
$WWZ$ vertices, the SM is assumed to be valid in the calculation. In
particular, the couplings of the weak bosons to quarks and leptons
are assumed to have their SM values. Section~II briefly summarizes
the technical details of our calculation.

The results of our numerical simulations are presented in Sec.~III.
In contrast to the SM contributions to the $q\bar q\to
W^+ W^-$ helicity amplitudes, terms
associated with non-standard $WWV$ couplings grow with energy.
Distributions which reflect the high energy behavior of the helicity
amplitudes, such as, the invariant mass distribution, the transverse
momentum spectrum of the charged lepton pair, or the transverse momentum
distribution of the individual leptons, are therefore very sensitive to
anomalous $WWV$ couplings. We identify the transverse momentum
distribution of the charged lepton pair, $d\sigma/dp_T(\ell_1^+\ell_2^-)$,
as the distribution which, at leading order (LO), is most sensitive to the
$WWV$ couplings, and discuss the impact of QCD corrections on this and
other distributions. In contrast to other distributions, the LO
$p_T(\ell_1^+\ell_2^-)$ distribution is not only sensitive to the high
energy behaviour of the $W^+W^-$ production amplitudes, but
also provides indirect information on
the helicities of the $W$ bosons, which are strongly correlated in $W$
pair production in the SM~\cite{LAGRANGIAN,BS,HEL}. Since anomalous $WWV$
couplings modify both the high energy behaviour of the amplitudes and
the correlations between the $W$ helicities,
$d\sigma/dp_T(\ell_1^+\ell_2^-)$ is particularly sensitive to these
couplings. We also
investigate in detail the background processes contributing to
$p\,p\hskip-7pt\hbox{$^{^{(\!-\!)}}$} \to W^+W^-+X\to \ell_1^+\nu_1
\ell_2^-\bar\nu_2+X$, in particular, the $t\bar t$ background. Both the
QCD corrections and the top quark background are found to be large.
They change the shape of the
$p_T(\ell_1^+\ell_2^-)$ distribution, and reduce the sensitivity to
anomalous $WWV$ couplings significantly.

In Sec.~III, we also show that
the size of the QCD corrections and the $t\bar t$ background can be
greatly reduced, and a significant fraction of the sensitivity lost can
be regained, if either a jet veto, or a cut on the transverse momentum
of the hadrons in the event, is imposed. Finally, we derive sensitivity limits
for anomalous $WWV$ couplings for various integrated
luminosities at the Tevatron and LHC, and compare them with those which can be
achieved in $W^\pm\gamma$ and $W^\pm Z$ production, and in $e^+e^-\to
W^+W^-$. Our conclusions are given in Sec.~IV.

\section{CALCULATIONAL TOOLS}

The calculation presented here
generalizes the results of Ref.~\cite{NEWJO} to include
arbitrary $C$ and $P$ conserving $WW\gamma$ and $WWZ$ couplings, and
employs a combination of analytic and Monte Carlo integration techniques.
Details of the method can be found in Ref.~\cite{NLOMC}.
The calculation is performed using the narrow width approximation for
the leptonically decaying $W$ bosons. In this approximation
difficulties in implementing finite $W$ width effects while
maintaining electromagnetic gauge invariance~\cite{GAUGE} are
automatically avoided, and it is straightforward to extend
the NLO calculation of $W^+ W^-$ production for on-shell $W$ bosons
to include the leptonic decays of the $W$ bosons.
Furthermore, non-resonant Feynman diagrams, such as
$u \bar u \to Z^{*} \to e^+ e^- Z$ followed by $Z \to \nu \bar \nu$,
contribute negligibly in this limit and can be ignored. Finite $W$ width
effects and non-resonant diagrams play an important role in the
$W$ pair threshold region. For the cuts we impose (see Sec.~IIIB), the
threshold region contributes negligibly to the cross section.

\subsection{Summary of ${\cal O}(\alpha_s)$ $W^+ W^-$ production including
leptonic $W$ decays}

The NLO calculation of $W^+ W^-$ production includes contributions from the
square of the Born graphs, the interference
between the Born graphs and the virtual one-loop diagrams, and the square
of the real emission graphs. The basic idea of the method employed here
is to isolate the
soft and collinear singularities associated with the real emission
subprocesses by partitioning phase space into soft, collinear, and
finite regions.  This is done by introducing theoretical soft and
collinear cutoff parameters, $\delta_s$ and $\delta_c$.  Using
dimensional regularization~\cite{DIMREG}, the soft and collinear
singularities are exposed as poles in  $\epsilon$ (the number of
space-time dimensions is $N = 4 - 2\epsilon$ with $\epsilon$ a small
number). The infrared singularities from the soft and virtual
contributions are then explicitly canceled while the collinear
singularities are factorized and absorbed into the definition of the
parton distribution functions. The remaining contributions are finite
and can be  evaluated in four dimensions.  The Monte Carlo program thus
generates $n$-body (for the Born and virtual contributions) and
$(n+1)$-body (for the real emission contributions) final state events.
The $n$- and $(n+1)$-body contributions both depend on the cutoff
parameters $\delta_s$ and $\delta_c$, however, when these contributions
are added together to form a suitably inclusive observable, all
dependence on the cutoff parameters cancels.
The numerical results presented in this paper are insensitive to
variations of the cutoff parameters.

Except for the virtual contribution,
the ${\cal O}(\alpha_s)$ corrections are all proportional to the Born cross
section.  It is easy to incorporate the leptonic $W$ decays
into those terms which
are proportional to the Born cross section; one simply replaces
$d\hat\sigma^{\hbox{\scriptsize Born}} (q \bar q \to W^+ W^-)$ with
$d\hat\sigma^{\hbox{\scriptsize Born}}
(q \bar q \to W^+ W^- \to \ell_1^+ \nu_1 \ell_2^- \bar\nu_2)$ in the relevant
formulae. When working at the amplitude level, the $W$ boson decays
are trivial to implement; the $W$ boson polarization vectors,
$\epsilon_\mu (k)$, are simply replaced by the corresponding
$W \to \ell \nu$ decay currents, $J_\mu (k)$, in the
amplitude.  Details of the amplitude level calculations for the Born
and real emission subprocesses can be found in Ref.~\cite{VVJET}.

The only term in which it is more difficult to incorporate the $W$
boson decays is
the virtual contribution.  Rather than undertake the non-trivial task of
recalculating the virtual correction term for the case of
leptonically decaying  $W$ bosons, we have instead opted to use the
virtual correction for real on-shell
$W$ bosons which we subsequently decay ignoring spin correlations.
When spin correlations are ignored, the spin summed squared matrix element
factorizes into separate production and decay squared matrix elements.
Neglecting spin correlations slightly modifies the shapes of the
angular distributions of the final state leptons, but
does not alter the total cross section as long as
no angular cuts ({\it e.g.}, rapidity cuts)
are imposed on the final state leptons. For realistic
rapidity cuts, cross sections are changed by typically 10\% when spin
correlations are neglected.
Since the size of the finite virtual correction is less than
$\sim 10\%$ the size of the Born cross section, the overall effect of
neglecting the spin correlations in the finite virtual correction is expected
to be negligible compared to the combined $10 - 20\%$ uncertainty
from the parton distribution functions, the choice of the factorization
scale $Q^2$, and higher order QCD corrections.

\subsection{Incorporation of Anomalous $WW\gamma$ and $WWZ$ Couplings}

The $WW\gamma$ and $WWZ$ vertices are uniquely determined in the SM by
$\hbox{\rm SU(2)} \bigotimes \hbox{\rm U(1)}$
gauge invariance.  In $W^+ W^-$ production the $W$ bosons
couple to essentially massless fermions, which
insures that effectively $\partial_\mu W^\mu=0$. This condition,
together with Lorentz invariance and conservation of $C$ and $P$,
allows six free parameters, $g_1^V, \kappa_V$, and $\lambda_V$
in the $WWV$ vertices ($V=\gamma,\,Z$).
The most general $WWV$ vertex, which is Lorentz, $C$, and $P$ invariant,
is described by the effective Lagrangian~\cite{LAGRANGIAN}
\begin{eqnarray}
\noalign{\vskip 5pt}
{\cal L}_{WWV} &=& -i \, g_{WWV}^{} \,
\Biggl[ g_1^V \bigl( W_{\mu\nu}^{\dagger} W^{\mu} V^{\nu}
                  -W_{\mu}^{\dagger} V_{\nu} W^{\mu\nu} \bigr)
+ \kappa_V W_{\mu}^{\dagger} W_{\nu} V^{\mu\nu}
+ {\lambda_V \over M_W^2} W_{\lambda \mu}^{\dagger} W^{\mu}_{\nu}
V^{\nu\lambda} \Biggr] \>,
\label{EQ:LAGRANGE}
\end{eqnarray}
\narrowtext
where $g_{WWV}^{}$ is the $WWV$ coupling strength ($g_{WW\gamma}^{} = e$ and
$g_{WWZ}^{} = e \cot\theta_{\rm W}$, where $e$ is the electric charge of
the proton and $\theta_{\rm W}$ is the weak mixing angle),
$W^{\mu}$ is the $W^-$ field, $V^{\mu}$ denotes the $Z$ boson
or photon field,
$W_{\mu\nu} = \partial_{\mu}W_{\nu} - \partial_{\nu}W_{\mu}$, and
$V_{\mu\nu} = \partial_{\mu}V_{\nu} - \partial_{\nu}V_{\mu}$.
At tree level in the SM, $g_1^V=1$, $\kappa_V = 1$, and $\lambda_V = 0$.
All higher dimensional operators are obtained by replacing $X^\mu$ with
$(\partial^2)^m X^\mu$ ($X=W,\,Z,\,\gamma$),
where $m$ is an arbitrary positive
integer, in the terms proportional to $\Delta g_1^V = g_1^V -1$,
$\Delta\kappa_V=\kappa_V-1$, and $\lambda_V$.
These operators form a complete set and can be summed by
replacing $\Delta g_1^V$, $\Delta\kappa_V$,
and $\lambda_V$ with momentum dependent form factors. All
details are contained in the specific functional form of the form factor
and its scale $\Lambda_{FF}$. For the $WW\gamma$ vertex,
electromagnetic gauge invariance requires that for on-shell photons
$\Delta g_1^\gamma = 0$. The corresponding form factor must hence be
proportional to some positive power of the square of the photon
momentum, $q^2_\gamma$. $\Delta g_1^\gamma$ therefore is of ${\cal
O}(q^2_\gamma/\Lambda_{FF}^2)$ and terms proportional to $\Delta g_1^\gamma$
in the helicity amplitudes are suppressed for momentum transfer smaller
than the form factor scale. To simplify our discussion somewhat, we
assume in the following that $\Delta g_1^\gamma=0$. The high energy
behavior of the form factors $\Delta g_1^Z$, $\Delta\kappa_V$, and
$\lambda_V$ will be discussed in more detail later in this section.

Following the standard notation of Ref.~\cite{LAGRANGIAN}, we have
chosen, without loss of generality, the
$W$ boson mass, $M_W$, as the energy scale in the denominator of the term
proportional to $\lambda_V$ in Eq.~(\ref{EQ:LAGRANGE}). If a different
mass scale, ${\tt M}$, had been used, then all of our subsequent
results could be obtained by scaling $\lambda_V$ by a factor
${\tt M}^2/M_W^2$.

At present, the $WWV$ coupling constants are only weakly constrained
experimentally (for a recent summary and discussion see
Ref.~\cite{DPF}). From a search performed in the channels $p\bar
p\to W^+W^-,\,W^\pm Z\to\ell^\pm\nu jj$ and $p\bar p\to
WZ \to jj \ell^+\ell^-$ ($\ell=e,\,\mu$) at large di-jet transverse momenta,
the CDF Collaboration obtains for $\Delta\kappa_\gamma=\Delta\kappa_Z$
and $\lambda_\gamma=\lambda_Z$~\cite{CDFWW}
\begin{eqnarray}
\noalign{\vskip 5pt}
-1.1<\Delta\kappa_V < 1.3~~({\rm for}~\lambda_V=\Delta g_1^V=0)
\>, \hskip 0.6cm
-0.8<\lambda_V < 0.8~~({\rm for}~\Delta\kappa_V=\Delta g_1^V=0) \>,
\label{EQ:LIM1}
\end{eqnarray}
at the 95\% confidence level (CL). Assuming that all other couplings take
their SM values, CDF also obtains a 95\% CL limit on $\Delta g_1^Z$ of
\begin{eqnarray}
\noalign{\vskip 5pt}
-1.2<\Delta g_1^Z < 1.2 \>.
\label{EQ:LIM2}
\end{eqnarray}
Slightly worse (better) limits on
$\Delta\kappa_\gamma$ ($\lambda_\gamma$) are obtained from $W^\pm\gamma$
production at the Tevatron~\cite{CDFWG,DOWG}.
{}From a comparison of their 95\% CL upper limit on the total
$W^+W^-\to\ell_1^+\nu_1\ell_2^-\bar\nu_2$ cross section
with the SM prediction, the D\O\ Collaboration finds for
$\Delta\kappa_\gamma=\Delta\kappa_Z$ and
$\lambda_\gamma=\lambda_Z$~\cite{DOWW}
\begin{eqnarray}
\noalign{\vskip 5pt}
-2.6<\Delta\kappa_V < 2.8~~({\rm for}~\lambda_V=\Delta g_1^V=0)
\>, \hskip 0.6cm
-2.2<\lambda_V < 2.2~~({\rm for}~\Delta\kappa_V=\Delta g_1^V=0) \>.
\label{EQ:LIM3}
\end{eqnarray}
To derive these limits, CDF (D\O) assumed a dipole form factor with scale
$\Lambda_{FF}=1.0$~TeV (0.9~TeV) [see below], however, the experimental
bounds are quite insensitive to the value of $\Lambda_{FF}$.

Although bounds on the $WWV$ couplings can also be extracted from
low energy data and oblique corrections to the 4-fermion $S$-matrix
elements, there are ambiguities and model-dependencies in the
results~\cite{DPF,DE,BL,HISZ,DV}.
{}From loop contributions to $(g-2)_\mu$~\cite{MUON}, $b\to
s\gamma$~\cite{BDEC,CLEO}, rare meson decays such as $K_L \to
\mu^+\mu^-$~\cite{HE} or $B\to K^{(*)}\mu^+\mu^-$~\cite{BAL},
$\epsilon'/\epsilon$~\cite{HMK}, and the $Z\to b\bar b$ width~\cite{NOV},
one estimates limits for the non-standard $WWV$ couplings of
${\cal O}(1-10)$. No rigorous bounds can be obtained from oblique
corrections, which combine~\cite{HMHK,ARZT} information from recent
LEP/SLD data, neutrino scattering experiments, atomic parity violation,
$\mu$-decay, and the $W$-mass measurement at hadron colliders,
if correlations between different
contributions to the anomalous couplings are fully taken into account.
Even without serious cancellations among various one loop contributions,
anomalous $WWV$ couplings of ${\cal O}(1)$ are still allowed by present
data~\cite{DPF,HISZ}. In contrast, invoking a ``naturalness'' argument
based on chiral perturbation theory~\cite{FLS,CHIRALP},
one expects deviations from the SM of ${\cal O}(10^{-2})$
or less for $g_1^V$, $\kappa_V$, and $\lambda_V$.

If $C$ or $P$ violating couplings are allowed, four additional
free parameters, $g_4^V$, $g_5^V$, $\tilde\kappa_V$ and
$\tilde\lambda_V$ appear in the effective $WWV$ Lagrangian~\cite{LAGRANGIAN}.
For simplicity, these couplings are not considered in this paper.

The Feynman rule for the $WWV$ vertex factor
corresponding to the Lagrangian in Eq.~(\ref{EQ:LAGRANGE}) is
\begin{eqnarray}
-i \, g_{WWV}^{} \, \Gamma_{\beta \mu \nu}^{} (k, k_1, k_2) =
-i \, g_{WWV}^{} \,
\biggl[
\Gamma_{\beta \mu \nu}^{\hbox{\scriptsize SM}} (k, k_1, k_2)
+ \Gamma_{\beta \mu \nu}^{\hbox{\scriptsize NSM}} (k, k_1, k_2) \biggr] \>,
\label{EQ:NSMCOUPLINGS}
\end{eqnarray}
where the labeling conventions for the four-momenta and Lorentz
indices are defined by Fig.~\ref{FIG:VERTEX}, and the factors
$\Gamma^{\hbox{\scriptsize SM}}$ and $\Gamma^{\hbox{\scriptsize NSM}}$
are the SM and non-standard model vertex factors:
\begin{eqnarray}
\noalign{\vskip 5pt}
\Gamma_{\beta \mu \nu}^{\hbox{\scriptsize SM}} (k, k_1, k_2) =
&\phantom{+}& (k_1 - k_2)_{\beta} \, g_{\nu \mu} + 2 \,
 k_{\mu} \, g_{\beta \nu} - 2 \, k_{\nu} \, g_{\beta \mu} \>, \\
\noalign{\vskip 5pt}
\Gamma_{\beta \mu \nu}^{\hbox{\scriptsize NSM}} (k, k_1, k_2)
= &\phantom{+}&
\left( \Delta g_1^V + \lambda_V \,{k^2 \over 2 M_W^2} \right)
        (k_1 - k_2)_{\beta} \, g_{\nu \mu} \\
\noalign{\vskip 5pt}
&-& {\lambda_V \over M_W^2} \, (k_1 - k_2)_{\beta} \, k_{\nu} \, k_{\mu}
 + (\Delta g_1^V + \Delta\kappa_V + \lambda_V) \, k_{\mu} \, g_{\beta \nu}
\nonumber \\
\noalign{\vskip 5pt}
&-& \left( \Delta g_1^V + \Delta\kappa_V + \lambda_V \right)
\, k_{\nu} \, g_{\beta \mu} \>. \nonumber
\end{eqnarray}
\narrowtext
The non-standard model vertex factor is written here in terms of
$\Delta g_1^V = g_1^V - 1$, $\Delta \kappa_V = \kappa_V - 1$,
and $\lambda_V$, which all vanish in the SM.

It is straightforward to include the non-standard model couplings in
the amplitude level calculations. The $q \bar q \to W^+ W^-$
virtual correction with the modified vertex factor of
Eq.~(\ref{EQ:NSMCOUPLINGS}) has been computed using the computer
algebra program FORM~\cite{FORM}, however, the resulting expression is
too lengthy to present here. The non-standard $WW\gamma$ and
$WWZ$ couplings of Eq.~(\ref{EQ:LAGRANGE}) do not
destroy the renormalizability of QCD. Thus the infrared singularities
from the soft and virtual contributions are explicitly canceled, and the
collinear singularities are factorized and absorbed into the definition
of the parton distribution functions, exactly as in the SM case.

The anomalous couplings can not be simply inserted into the vertex
factor as constants because this would violate $S$-matrix unitarity.
Tree level unitarity uniquely restricts the $WWV$ couplings to their
SM gauge theory values at asymptotically high energies~\cite{CORNWALL}.
This implies that any deviation of $\Delta g_1^V$, $\Delta \kappa_V$,
or $\lambda_V$ from the SM expectation has to be described by a form
factor $\Delta g_1^V(M_{WW}^2, p^2_{W^+}, p^2_{W^-})$,
$\Delta \kappa_V(M_{WW}^2, p^2_{W^+}, p^2_{W^-})$, or
$\lambda_V(M_{WW}^2, p^2_{W^+}, p^2_{W^-})$
which vanishes when either the square of the $W^+ W^-$ invariant mass,
$M_{WW}^2$, or the
square of the four-momentum of a final state $W$ boson ($p^2_{W^+}$ or
$p^2_{W^-})$ becomes large.  In $W^+ W^-$ production $p_W^2 \approx M_W^2$
even when the finite $W$ width is taken into account.
However, large values of $M_{WW}^2$ will be probed at future hadron
colliders like the LHC and the $M_{WW}^2$ dependence
of the anomalous couplings has
to be included in order to avoid unphysical results which would violate
unitarity.  Consequently, the anomalous couplings (denoted generically by $a$,
$a = \Delta g_1^V, \Delta \kappa_V, \lambda_V$)
are introduced via form factors~\cite{FORMF}. The functional behaviour
of the form factors depends on the details of the underlying new
physics. Effective Lagrangian techniques are of little help here because
the low energy expansion which leads to the effective Lagrangian exactly
breaks down where the form factor effects become important. Therefore,
ad hoc assumptions have to be made. Here, we assume a behaviour similar
to the nucleon form factor
\begin{eqnarray}
a(M_{WW}^2, p^2_{W^+} = M_W^2, p^2_{W^-} = M_W^2) \> &=& \>
{a^0 \over (1 + M_{WW}^2/\Lambda_{FF}^2)^n } \>,
\label{EQ:AFORM}
\end{eqnarray}
where $a^0$ is the form factor value at low energies and
$\Lambda_{FF}$ represents the scale at which new physics becomes
important in the weak boson sector.
In order to guarantee unitarity, it is necessary to have $n>1$.
For the numerical results presented here, we use a dipole
form factor ($n=2$) with a scale $\Lambda_{FF} = 1$~TeV, unless
explicitly stated otherwise. The exponent $n=2$
is chosen in order to suppress $W^+ W^-$ production at energies
$\sqrt{\hat s} > \Lambda_{FF} \gg M_W$, where novel phenomena
like resonance
or multiple weak boson production are expected to become important.

Form factors are usually not introduced if an ansatz based on chiral
perturbation theory is used.  In the framework of chiral perturbation
theory, the effective Lagrangian describing the anomalous vector boson
self-interactions breaks down at center of mass energies above a few
TeV~\cite{FLS,CHIRALP} (typically $4\pi v \sim 3$~TeV, where $v\approx
246$~GeV is the Higgs field vacuum expectation value).
Consequently, one has to limit the center of mass energies to
values sufficiently below $4\pi v$ in this approach.

The electroweak symmetry can either be realized in a
linear~\cite{DPF,HISZ} or non-linear way~\cite{DPF,BL,DV}. If the
$\hbox{\rm SU(2)} \bigotimes \hbox{\rm U(1)}$
symmetry is realized linearly, and only dimension~6 operators are considered,
there are 11 independent, $\hbox{\rm SU(2)} \bigotimes \hbox{\rm
U(1)}$ invariant, dimension~6 operators~\cite{BUCH}. Three of these
operators give rise to non-standard $WWV$ couplings~\cite{HISZ}. In this
scenario, both anomalous $WW\gamma$ and $WWZ$ couplings are
simultaneously non-zero. Assuming, for simplicity, that
the coefficients of the two operators which generate non-zero values of
$\Delta\kappa_\gamma$ and $\Delta\kappa_Z$ are equal, only two independent
anomalous couplings remain
(this scenario is known as the Hagiwara-Ishihara-Szalapski-Zeppenfeld (HISZ)
scenario [see Ref.~\cite{HISZ}]). Choosing, for example,
$\Delta\kappa_\gamma$ and $\lambda_\gamma$ as independent parameters,
the $WWZ$ couplings are then given by:
\begin{eqnarray}
\Delta g_1^Z & = & {1\over 2\cos^2\theta_W}\,\Delta\kappa_\gamma ,
\label{EQ:HISZ1} \\[1.mm]
\Delta\kappa_Z & = & {1\over 2}\,(1-\tan^2\theta_W)\,\Delta\kappa_\gamma
, \label{EQ:HISZ2}
\\[1.mm]
\lambda_Z & = & \lambda_\gamma .
\label{EQ:HISZ3}
\end{eqnarray}
In Secs.~IIIE and~IIIG we shall use the HISZ scenario, defined by these
equations,
as a simple and illustrative example of a model where both $WW\gamma$
and $WWZ$ couplings simultaneously deviate from their SM values.
Equations~(\ref{EQ:HISZ1}) --~(\ref{EQ:HISZ3}) are modified
when operators of dimension~8 or higher are
incorporated~\cite{HISZ}, which may introduce large
corrections~\cite{ARZT}. Different relations are obtained by invoking
global symmetry arguments, or by fine tuning anomalous $WWV$ couplings
such that the most serious unitarity violating
contributions to the tree level vector boson scattering amplitudes are
avoided~\cite{BIL}.

\section{PHENOMENOLOGICAL RESULTS}

We shall now discuss the phenomenological implications of ${\cal
O}(\alpha_s)$ QCD corrections in $W^+ W^-$ production at the Tevatron
($p\bar p$ collisions at $\sqrt{s} = 1.8$~TeV) and the LHC ($pp$
collisions at $\sqrt{s} = 14$~TeV). We first briefly describe
the input parameters, cuts, and the finite energy resolution smearing
used to simulate detector response. We then explore the sensitivity
of the observables in $W^+W^-\to\ell_1^+\nu_1\ell^-_2\bar\nu_2$ to
anomalous $WWV$ couplings, and discuss in detail the impact
of ${\cal O}(\alpha_s)$ QCD corrections and various background processes
on the observability of non-standard $WWV$
couplings in $W^+ W^-$ production at the Tevatron and LHC. To simplify
the discussion, we shall concentrate on the channel $W^+W^-\to e^+\nu_e
e^-\bar\nu_e$. In absence of lepton flavor
specific cuts, the cross sections for $W^+W^-\to e^+\nu_e
e^-\bar\nu_e$ and the other three leptonic channels,
$W^+W^-\to\mu^+\nu_\mu\mu^-\bar\nu_\mu$, $W^+W^-\to\mu^+\nu_\mu
e^-\bar\nu_e$ and $W^+W^-\to e^+\nu_e\mu^-\bar\nu_\mu$ are equal.
Decay modes where one or both charged leptons in the final state
originate from $W\to\tau\nu_\tau\to e\nu_e\bar\nu_\tau\nu_\tau$ are
discussed in Sec.~IIIF. No attempt is made to include the contributions
from gluon fusion, $gg\to W^+W^-$, into our calculations, which formally
are of ${\cal O}(\alpha_s^2)$. Gluon fusion contributes less than 1\%
(15\%) to the total $W$ pair cross section at the Tevatron
(LHC)~\cite{KAO}.

\subsection{Input Parameters}

The numerical results presented here
were obtained using the two-loop expression for
$\alpha_s$. The QCD scale $\Lambda_{\hbox{\scriptsize QCD}}$
is specified for four
flavors of quarks by the choice of the parton distribution functions  and
is adjusted whenever a heavy quark threshold is crossed so that
$\alpha_s$ is a continuous function of $Q^2$. The heavy quark masses
were taken to be $m_b=5$~GeV and $m_t=176$~GeV~\cite{TOPMASS1,TOPMASS2}.

The SM parameters used in the numerical simulations are $M_Z = 91.19$~GeV,
$M_W = 80.22$~GeV, $\alpha (M_W) =1/128$, and $\sin^2
\theta_{\hbox{\scriptsize W}} = 1 - (M_W^{}/M_Z^{})^2$. These values are
consistent with recent measurements at LEP, SLC, the CERN $p\bar p$
collider, and the Tevatron~\cite{LEP,SLC,MW}. The soft and collinear
cutoff parameters, discussed in Sec.~IIA,
are fixed to $\delta_s = 10^{-2}$ and $\delta_c = 10^{-3}$. The parton
subprocesses have been summed over $u,d,s$, and $c$ quarks.
The $W$ boson leptonic branching ratio is taken to be
$B(W \to \ell \nu) = 0.107$ and the total width of the $W$ boson is
$\Gamma_W = 2.08$~GeV. Except where otherwise stated, a single scale
$Q^2=M^2_{WW}$,  where $M_{WW}$ is the invariant mass of the
$W^+ W^-$ pair, has been used for the renormalization scale $\mu^2$
and the factorization scale $M^2$.
The NLO numerical results have been calculated in the modified Minimal
Subtraction ($\overline {\rm MS}$) scheme~\cite{MSBAR}.

In order to get consistent NLO results it is necessary to use parton
distribution functions which have been fit to next-to-leading order.
Our numerical simulations have been performed using
the Martin-Roberts-Stirling (MRS)~\cite{MRSA} set~A distributions
($\Lambda_4 = 230$~MeV)
in the $\overline {\rm MS}$ scheme. They take into account recent
measurements of the proton structure functions at HERA~\cite{HERA}, the
asymmetry of the rapidity distribution of the charged lepton from
$W^\pm\to\ell^\pm\nu$~\cite{ASYMM}, and the asymmetry in Drell-Yan
production in $pp$ and $pn$ collisions~\cite{NA51}. For convenience,
the MRS set~A distributions have also been used for the LO calculations.

\subsection{Cuts}

The cuts imposed in the numerical simulations are motivated by
the finite acceptance of the detectors.
The complete set of transverse momentum ($p_T^{}$) and pseudorapidity ($\eta$)
cuts can be summarized as follows.
\begin{quasitable}
\begin{tabular}{cc}
Tevatron & LHC\\
\tableline
$p_{T}^{}(e)         > 20$~GeV  & $p_{T}^{}(e)         > 25$~GeV\\
$p\llap/_T^{}           > 30$~GeV  & $p\llap/_T^{}           > 50$~GeV\\
$|\eta(e)|           < 2.5$     & $|\eta(e)|           < 3.0$\\
\end{tabular}
\end{quasitable}
The large missing transverse momentum ($p\llap/_T^{}$) cut  has been
chosen to reduce potentially dangerous backgrounds from event
pileup~\cite{PILE}
and processes where particles outside the rapidity range covered by the
detector contribute to the missing transverse momentum. These
backgrounds are potentially dangerous at the LHC with its large design
luminosity of ${\cal L}=10^{34}$~cm$^{-2}$~s$^{-1}$~\cite{LHC}, and also
the TeV* under certain conditions. In several of the TeV* scenarios
which are currently under investigation~\cite{Brenna,GPJ}, the average number
of interactions
per bunch crossing is similar to that expected at the LHC. Present
studies for the LHC~\cite{ATLAS,CMS} and extrapolations to Tevatron
energies indicate that these backgrounds
are under control for the $p\llap/_T^{}$ cuts listed above. The total
$W^+ W^-\to e^+\nu_e e^-\bar\nu_e$, cross section within cuts in the
Born approximation at the Tevatron and LHC is 0.04~pb and
0.15~pb, respectively.

\subsection{Finite Energy Resolution Effects}

Uncertainties in the energy measurements of the charged leptons
in the detector are simulated in the calculation by Gaussian smearing
of the particle four-momentum vector with standard deviation $\sigma$.
For distributions which require a jet definition, {\it e.g.}, the $W^+ W^- +
1$~jet exclusive cross section, the jet four-momentum vector is also
smeared. The standard deviation $\sigma$
depends on the particle type and the detector. The numerical results
presented here for the Tevatron and LHC center of mass energies
were made using $\sigma$ values based on the CDF~\cite{RCDF} and
ATLAS~\cite{ATLAS} specifications, respectively.

\subsection{Signatures of Anomalous $WWV$ Couplings and ${\cal
O}(\alpha_s)$ Corrections}

In contrast to the SM contributions to the $q\bar q\to
W^+ W^-$ helicity amplitudes, terms
associated with non-standard $WWV$ couplings grow with energy.
A typical signal for anomalous couplings therefore will be a broad
increase in the invariant mass distribution of the $W$ pair at large
values of the invariant mass, $M_{WW}$. Due to the fact that
non-standard $WWV$ couplings only contribute via $s$-channel photon and
$Z$ exchange, their effects are concentrated in the region of small $W$
rapidities, and the $W$ transverse momentum distribution is particularly
sensitive to anomalous couplings. However, if both $W$ bosons decay
leptonically, $W^+W^-\to e^+\nu_e e^-\bar\nu_e$, the $W^+W^-$ invariant
mass and the $W$ transverse momentum cannot be reconstructed
since the two neutrinos are not observed.

Alternatively, the invariant mass distribution of the $e^+e^-$ pair, or
the electron or positron $p_T$ spectrum can be studied. The differential
cross section for $p_T(e)$ in the reaction $p\bar p\to W^+W^-+X\to
e^+e^-p\llap/_T+X$ at $\sqrt{s}=1.8$~TeV is shown in Fig.~\ref{FIG:TWO}.
The Born and NLO results are shown in Fig.~\ref{FIG:TWO}a and
Fig.~\ref{FIG:TWO}b, respectively. Both the $e^+$ and $e^-$
transverse momenta
are histogrammed, each with half the event weight.
Results are displayed
for the SM and for five sets of anomalous couplings, namely,
($\lambda^0_\gamma=-0.5$, $\Delta\kappa^0_\gamma=0$, SM $WWZ$ couplings),
($\Delta\kappa^0_\gamma=-0.5$, $\lambda^0_\gamma=0$, SM $WWZ$ couplings),
($\lambda^0_Z=-0.5$, $\Delta g_1^{Z0}=\Delta\kappa^0_Z=0$, SM
$WW\gamma$ couplings),
($\Delta\kappa^0_Z=-0.5$, $\Delta g_1^{Z0}=\lambda^0_Z=0$, SM
$WW\gamma$ couplings), and
($\Delta g_1^{Z0}=-1$, $\Delta\kappa^0_Z=\lambda^0_Z=0$, SM
$WW\gamma$ couplings). For simplicity, only one anomalous
coupling at a time is allowed to differ from its SM value.
The figure shows that at the Tevatron center of mass energy,
NLO QCD corrections do not have a large influence on the sensitivity of
the $p_T(e)$ distribution to anomalous couplings. The
${\cal O}(\alpha_s)$ corrections at Tevatron energies are approximately
30~--~40\% for the SM as well as for the anomalous coupling cases. Due
to the larger coupling of the $Z$ boson to quarks and $W$ bosons [see
Eq.~(\ref{EQ:LAGRANGE})], anomalous $WWZ$ couplings yield larger
differences from the SM than non-standard $WW\gamma$ couplings of the
same type and strength. Whereas terms proportional to
$\lambda_V$ and $\Delta\kappa_V$ in the helicity amplitudes grow like
$\hat s/M_W^2$, terms associated with $\Delta g_1^Z$ only increase with
$\sqrt{\hat s}/M_W$~\cite{LAGRANGIAN}. As a result, the sensitivity of
$W^+W^-$ production
to non-standard values of $g_1^Z$ is considerably smaller than
it is for $\Delta\kappa_V$ and $\lambda_V$.

For $\Delta\kappa^0_V$ ($\Delta g_1^{Z0}$), positive anomalous
couplings lead to $\sim 40\%$ ($\sim 20\%$) smaller deviations from the
SM prediction in the $p_T(e)$ distribution
than negative non-standard couplings of equal magnitude, whereas
the sign makes
little difference for $\lambda^0_V$. This statement also applies to
other distributions. This effect can be easily understood from the high
energy behaviour of the $W^+W^-$ production amplitudes, ${\cal
M}(\lambda_{W^+}, \lambda_{W^-})$, where $\lambda_{W^\pm}$ denotes the
helicity of the $W^\pm$ boson~\cite{LAGRANGIAN}. Any dependence of the
differential cross section on the sign of one of
the anomalous coupling parameters originates from interference effects
between the SM and the anomalous terms in the helicity amplitudes.
In the SM, only ${\cal M}(\pm,\mp)$ and ${\cal
M}(0,0)$ remain finite for $\hat s\to\infty$. Contributions to the
helicity amplitudes proportional to $\lambda_V$ mostly influence the
$(\pm,\pm)$ amplitudes. The SM ${\cal
M}(\pm,\pm)$ amplitudes vanish like $1/\hat s$, and the non-standard terms
dominate except for the threshold region, $\sqrt{\hat s}\approx
2M_W$. For non-standard values of $\lambda_V$, the cross section
therefore depends only very little on the sign of the anomalous
coupling. Terms proportional to $\Delta\kappa_V$ also increase like $\hat
s/M_W^2$ with energy, but mostly contribute to the (0,0) amplitude,
which remains finite in the SM in the high energy limit. Interference
effects between the SM and the anomalous contributions to the (0,0)
amplitude, thus, are non-negligible, resulting in a significant
dependence of the differential cross section on the sign of
$\Delta\kappa_V$.  Finally, terms
proportional to $\Delta g_1^Z$ are proportional to $\sqrt{\hat
s}/M_W$ and mostly influence the amplitudes with one longitudinal and
one transverse $W$ boson. In the SM, these terms vanish like $1/\sqrt{\hat
s}$. The dependence on the sign of $\Delta g_1^Z$ is, therefore, less
pronounced than for $\Delta\kappa_V$.

The $p_T(e)$ distribution at the LHC is shown in Fig.~\ref{FIG:THREE}.
At leading order, the sensitivity of the electron transverse momentum
distribution to anomalous $WWV$ couplings is significantly more
pronounced than at the Tevatron. Because of the form
factor parameters assumed, the result for $\Delta g_1^{Z0}=-1$ approaches
the SM result at large values of $p_T(e)$. As mentioned before, we have
used $n=2$ and a form factor scale of $\Lambda_{FF}=1$~TeV in all our
numerical simulations [see Eq.~(\ref{EQ:AFORM})].
For a larger scale $\Lambda_{FF}$, the deviations
from the SM result become more pronounced at high energies. In contrast
to the situation encountered at the Tevatron, the shape of the
SM $p_T(e)$ spectrum at the LHC is considerably affected by NLO QCD
corrections. At $p_T(e)=600$~GeV, the QCD corrections increase the SM
cross section by about a factor~4, whereas the enhancement is only a
factor~1.5 at $p_T(e)=100$~GeV. In the presence of anomalous
couplings, the higher order QCD corrections are much smaller than in the
SM. In regions where the anomalous terms dominate, the ${\cal
O}(\alpha_s)$ corrections are typically between 30\% and 40\%.
At next-to-leading order, the sensitivity of the electron transverse
momentum spectrum to anomalous couplings thus is considerably reduced at
the LHC.

The large QCD corrections at high values of $p_T(e)$ are caused by a
collinear enhancement factor, $\log^2(p_T(W)/M_W)$, in the
$qg\to W^+W^-q$ partonic cross
section for $W$ transverse momenta much larger than $M_W$,
$p_T(W)\gg M_W$, and the large $qg$ luminosity at LHC
energies~\cite{WWFRIX}.
It arises from the kinematical region where one of the $W$ bosons is
produced at large $p_T^{}$ and recoils against the quark, which radiates a
soft $W$ boson which is almost collinear to the quark, and thus is
similar in nature to the enhancement of QCD corrections observed at
large photon and $Z$ boson transverse momenta in $W\gamma$ and $WZ$
production~\cite{WVENHANC,BHO,BHO1}. Since the Feynman
diagrams contributing in the collinear approximation
do not involve the $WWV$ vertices, the logarithmic enhancement
factor only affects the SM matrix elements.

Although non-standard $WWV$ couplings lead to a large enhancement in the
differential cross section of the lepton transverse momentum in
$W^+W^-\to\ell_1^+\ell_2^-p\llap/_T$ production, the sensitivity is,
due to the phase space effect of the $W$ decays, significantly
reduced compared to that of the photon ($Z$) transverse momentum distribution
in $W\gamma$ ($WZ$) production~\cite{BHO,BHO1}.
As an alternative to the averaged charged lepton $p_T$ distribution, the
differential cross sections of the maximum and minimum lepton transverse
momenta can be studied. The distribution of the maximum lepton $p_T$
exhibits a sensitivity to non-standard $WWV$ couplings similar to that
encountered in the average lepton $p_T$ distribution. The minimum lepton
transverse momentum distribution, on the other hand, is very insensitive
to anomalous couplings.
In contrast to the charged lepton $p_T$ distribution, the shape of the
invariant mass spectrum of the $e^+e^-$ pair remains essentially
unaffected by QCD corrections. However, the $M(e^+e^-)$ distribution is
found to be considerably less sensitive to anomalous $WWV$ couplings
than the transverse momentum spectrum of the charged leptons. The
cluster transverse mass distribution exhibits a sensitivity to
non-standard $WWV$ couplings which is quite similar to that found in the
$p_T(e)$ distribution.

In Figs.~\ref{FIG:FOUR} and~\ref{FIG:FIVE} we show the differential
cross section for the missing transverse momentum. The leading order
$p\llap/_T$ spectrum is seen to be considerably more sensitive to non-standard
$WWV$ couplings than the $p_T(e)$ distribution. The relatively large
missing $p_T$ cut which we impose at both the Tevatron and the LHC does
not noticeably reduce the sensitivity to anomalous $WWV$ couplings.
QCD corrections strongly affect the shape of the $p\llap/_T$
distribution, and reduce the sensitivity to anomalous couplings. At the
LHC this effect is very dramatic (see Fig.~\ref{FIG:FIVE}); the NLO
missing $p_T$ spectrum is seen to be considerably less sensitive to
non-standard $WWV$ couplings than
the NLO $p_T(e)$ distribution (see Fig.~\ref{FIG:THREE}b).

The effect
of the QCD corrections is shown in more detail in Fig.~\ref{FIG:SIX},
where we display the ratio of the NLO and LO differential cross sections
for the missing transverse momentum and the $p_T$ of the charged
leptons. Both, at Tevatron and LHC energies, the ${\cal O}(\alpha_s)$
corrections are approximately 30\% at small $p\llap/_T$ values. The NLO
to LO differential cross section ratio
begins to rise rapidly for $p\llap/_T> 70$~GeV, and for
$p\llap/_T=200$~GeV (600~GeV) the QCD corrections increase the cross
section by a factor $\sim 5$ ($\sim 100$) at the Tevatron (LHC). The shape
change in the $p\llap/_T$ distribution thus is much more pronounced than
that observed in the charged lepton transverse momentum distribution.

In the SM, the dominant $W^\pm$ helicity at high energies in
$\bar uu\to W^+W^-$ ($\bar dd\to W^+W^-$) is $\lambda_{W^\pm}=\mp 1$
($\lambda_{W^\pm}=\pm 1$)~\cite{LAGRANGIAN,BS,HEL} because of a
$t$-channel pole factor which peaks at small scattering angles with an
enhancement factor which is proportional to $\hat s$. Due to the $V-A$
nature of the $We\nu$ coupling, the angular
distribution of the neutrino in the rest frame of the parent $W$ is
proportional to $(1-Q_W\lambda_W\cos\theta)^2$, where $Q_W$ is the
$W$ charge and $\theta$ is the
angle with respect to the flight direction of the $W$ in the parton
center of mass frame. As a result, the neutrinos tend
to be emitted either both into ($\bar uu$ annihilation), or both
against the flight direction of their parent $W$ boson ($\bar dd$
annihilation), {\it i.e.}, they reflect the kinematical properties of the
$W$ bosons. At leading order, the $W^+$ and the $W^-$ in $W$ pair
production are back to back in the transverse plane, and the
transverse momenta  of the two neutrinos tend to cancel at high
energies. Above the $W$ threshold, the SM missing transverse momentum
distribution thus drops much more rapidly than the $p_T$ distribution of
the charged leptons.

Anomalous $WWV$ couplings tend to destroy the correlation of the
neutrino momenta. Non-standard values of $\Delta\kappa_V$ mostly contribute to
the amplitude where both $W$'s are longitudinal. Terms in the helicity
amplitudes proportional to $\Delta g_1^Z$ predominantly affect the $(0,
\pm)$ and $(\pm,0)$ amplitudes, and non-zero values of $\lambda_V$
mostly contribute to $(\pm,\pm)$ states, with equal numbers of $W$'s of
positive and negative helicity~\cite{LAGRANGIAN}. The angular distribution of
the $W$ decay lepton for a longitudinal $W$ boson is proportional to
$\sin^2\theta$, whereas equal numbers of $W$'s with $\lambda_W=+1$ and
$\lambda_W=-1$ produce a $(1+\cos^2\theta)$ spectrum. As a result, the
cancellation of the transverse momenta of the neutrinos is less perfect
in the presence of anomalous couplings. This reinforces the growth of the
non-standard contributions to the helicity amplitudes with energy, thus
producing a very pronounced sensitivity of the LO $p\llap/_T$
distribution to anomalous $WWV$ couplings.

The delicate balance of the neutrino transverse momenta, however, is
also spoiled by the real emission processes ($q\bar q\to W^+W^-g$ {\it
etc.}) which contribute to the ${\cal O}(\alpha_s)$ QCD corrections. At
large transverse momenta, QCD corrections therefore affect the
$p\llap/_T$ distribution much more than the charged lepton $p_T$
spectrum; see Fig.~\ref{FIG:SIX}.

Experimentally, the missing transverse momentum distribution is more
difficult to measure than other differential cross sections due to
cracks and other detector imperfections which give rise to ``fake''
$p\llap/_T$, or worsen the resolution of the missing $p_T$ distribution.
At lowest order, the $p\llap/_T$ vector is balanced by the transverse
momentum vector of the charged lepton pair,
which we denote by $p_T(e^+ e^-)$
[${\bf p}_T(e^+ e^-) \equiv {\bf p}_T(e^+) + {\bf p}_T(e^-)$].
The angular distribution of the charged leptons in the rest frame of
the parent $W$ can be obtained from that of the neutrino by replacing
the angle $\theta$ by $\pi+\theta$. As a result, the charged
lepton transverse momentum vectors are also strongly correlated.
The $p_T(e^+e^-)$ differential cross section, which can readily be measured
experimentally, is therefore expected to exhibit a sensitivity to
anomalous $WWV$ couplings and ${\cal O}(\alpha_s)$ QCD corrections
similar to that of the $p\llap/_T$ distribution. The transverse momentum
distribution of the charged lepton pair at the Tevatron and LHC is shown
in Figs.~\ref{FIG:SEVEN} and~\ref{FIG:EIGHT}, respectively. At high
values of $p_T(e^+e^-)$, the transverse momentum spectrum of the charged
lepton pair is seen to be very similar to the $p\llap/_T$ distribution,
with a similar sensitivity to anomalous $WWV$ couplings and to ${\cal
O}(\alpha_s)$ QCD corrections. At small values, the LO $p_T(e^+e^-)$
and $p\llap/_T$ distributions differ due to the smearing imposed on the
charged lepton momenta.

{}From the picture outlined above, one expects that at next-to-leading
order, $W^+W^-$ events with a large missing transverse momentum or a
high $p_T$ charged lepton pair, will most of the time contain a high
transverse momentum jet.
This fact is illustrated in Fig.~\ref{FIG:NINE} which shows
the decomposition
of the inclusive SM NLO $p_T(e^+e^-)$ differential cross section into NLO
0-jet and LO 1-jet exclusive cross sections at the Tevatron
and LHC.
For comparison, the
$p_T(e^+e^-)$ distribution obtained in the Born approximation is also
shown in the figure.
Here, a jet is defined as a quark or gluon with
\begin{eqnarray}
p_T^{}(j)>20~{\rm GeV}\hskip 1.cm {\rm and} \hskip 1.cm |\eta(j)|<2.5
\label{EQ:TEVJET}
\end{eqnarray}
at the Tevatron, and
\begin{eqnarray}
p_T^{}(j)>50~{\rm GeV}\hskip 1.cm {\rm and} \hskip 1.cm |\eta(j)|<3
\label{EQ:LHCJET}
\end{eqnarray}
at the LHC. The sum of the NLO 0-jet and the LO 1-jet exclusive cross
section is equal to the inclusive NLO cross section.
The results for the NLO exclusive $W^+W^-+0$~jet and the LO exclusive
$W^+W^-+1$~jet differential cross sections depend explicitly on the jet
definition. Only the inclusive NLO distributions are independent of the
jet definition.

Present LHC studies~\cite{ATLAS,CMS,LHCJET} and projections to Tevatron
energies suggest that jets fulfilling the
criteria of Eqs.~(\ref{EQ:TEVJET}) and~(\ref{EQ:LHCJET}) can be
identified without problems at the TeV*~\cite{GPJ} and LHC~\cite{LHC}
design luminosities of $10^{33}$~cm$^{-2}$~s$^{-1}$ and
$10^{34}$~cm$^{-2}$~s$^{-1}$, respectively. For luminosities
significantly below the design luminosity, it may well be possible to
lower the jet-defining $p_T$ threshold to 10~GeV at the Tevatron and
30~GeV at the LHC. It should be noted, however, that for theoretical
reasons, the jet transverse momentum threshold can not be made
arbitrarily small in our calculation. For transverse momenta below
5~GeV (20~GeV) at the Tevatron (LHC), soft gluon resummation effects
are expected to significantly change the shape of the jet $p_T^{}$
distribution~\cite{RESUM}. For the jet definitions discussed above,
these effects are expected to be unimportant and therefore are ignored
in our calculation.

Figure~\ref{FIG:NINE} shows that, at the Tevatron, the 1-jet cross
section is larger than
the 0-jet rate for $p_T(e^+e^-)>100$~GeV, and dominates completely at
large $p_T(e^+e^-)$. The NLO 0-jet and Born differential cross sections
deviate by at most 30\% for lepton pair transverse momenta above 30~GeV
(60~GeV) at the Tevatron (LHC). For $p_T(e^+e^-)<25$~GeV (40~GeV) at the
Tevatron (LHC), the 1-jet cross section again dominates. In this region
the 0-jet cross section is strongly suppressed due to the cut imposed on
the missing transverse momentum. Figure~\ref{FIG:NINE} suggests that the
size of the QCD corrections in the $p_T(e^+e^-)$ distribution can be
dramatically reduced by vetoing hard jets in the central rapidity
region, {\it i.e.}, by imposing a ``zero jet'' requirement and
considering the $W^+W^-+0$~jet channel only.

As mentioned in Sec.~IIIA, all our results are obtained for
$Q^2=M^2_{WW}$. The Born cross section for $W$ pair production depends
significantly on the choice of $Q$, which enters through the
scale-dependence of the parton distribution functions. At the NLO level,
the  $Q$-dependence enters not only via the
parton distribution functions, but also through the running coupling
$\alpha_s(Q^2)$ and the explicit factorization scale-dependence in the
order $\alpha_s(Q^2)$ correction terms. Similar to the situation
encountered in $W\gamma$ and $WZ$ production in hadronic
collisions~\cite{BHO,BHO1}, we find that the NLO $W^+W^-+0$~jet exclusive
cross section is almost independent of the scale $Q$. Here,
the scale-dependence of the
parton distribution functions is compensated by that of $\alpha_s(Q^2)$
and the explicit factorization scale dependence in the correction terms.
The $Q$-dependence of the inclusive NLO cross section is significantly
larger than that of the NLO 0-jet cross section; it is dominated by the
1-jet exclusive component which is calculated only to lowest order and
thus exhibits a considerable scale-dependence.

\subsection{Background Processes}

So far, we have only considered the $W^+W^-\to e^+e^-p\llap/_T+X$ signal
cross section. However, a number of processes lead to the same final
states. These processes contribute to the background and,
in addition to the NLO QCD corrections, reduce the sensitivity to
anomalous $WWV$ couplings. The situation is summarized in
Fig.~\ref{FIG:TEN}, where we show, at leading order, the transverse
momentum distribution of the charged lepton pair for the $W^+W^-$ signal
(solid lines), and the most important background processes.

The potentially most dangerous background originates from top quark pair
production, $p\,p\hskip-7pt\hbox{$^{^{(\!-\!)}}$} \to t\bar t\to
W^+W^-b\bar b\to e^+e^-p\llap/_T+X$. To compute the top quark production
rate, we use the matrix elements of the full processes $q\bar q,~gg
\to t \bar t \to W^+W^-b\bar b\to f_1\bar f_2f_3\bar f_4b\bar b$~\cite{KS}.
We assume
that the SM correctly describes the production and decay of top quarks.
At present, the mass and the properties of the top quark are still
rather poorly known, and although the SM predictions are in agreement
with the experimental data~\cite{TOPMASS1,TOPMASS2}, there is
substantial room for non-SM physics. The CDF
Collaboration obtains its most precise measurement of the top quark
mass, $m_t=176\pm 8\pm
10$~GeV, from a sample of $b$-tagged $W+$~jets events~\cite{TOPMASS1},
whereas D\O\ finds $m_t=199^{+19}_{-21}\pm 22$~GeV~\cite{TOPMASS2} from
a combined analysis of all available channels. In the following, for
definiteness, we take $m_t=176$~GeV. For larger values of $m_t$, the
top quark background is reduced; the $t\bar t$ cross section drops by about a
factor~2 (1.7) at the Tevatron (LHC) if the top quark mass is increased
to 200~GeV.

For the cuts we impose (see
Sec.~IIIA), the $W^+W^-$ and $t\bar t$ total cross sections are approximately
equal at the Tevatron. However, due to the $b$-quarks produced in the
decay of the $t$ and $\bar t$, the $p_T(e^+e^-)$ distribution from $t\bar
t$ production is considerably broader and harder than that of the
charged lepton pair in $W^+W^-$ production. At large values of
$p_T(e^+e^-)$, the top quark background (dashed line) therefore
completely
dominates over the $W$ pair signal at the Tevatron. At the LHC, the
$t\bar t$ cross section is approximately a factor~25 larger than the
$W^+W^-$ rate, and the top quark background is at least a factor~10
bigger than the signal over the entire range of lepton pair transverse
momenta (see Fig.~\ref{FIG:TEN}b). For $m_t=200$~GeV, the $p_T(e^+e^-)$
differential cross section almost coincides with that obtained for
$m_t=176$~GeV for $p_T(e^+e^-)>150$~GeV; only for smaller values of the
lepton pair transverse momentum does the larger mass reduce the rate.

$W^\pm Z$ production where both the $W$ and the $Z$ boson decay
leptonically may also contribute to the background if one of the two
like sign charged leptons is produced with a rapidity outside the range
covered by the detector. To estimate the $W^\pm Z$ background, we have
assumed that, at the Tevatron (LHC), charged leptons with $p_T(\ell)<
10$~GeV (15~GeV) or $|\eta(\ell)|>2.5$ (3.0) are not detected, and thus
contribute to the missing transverse momentum vector. Our results, represented
by the long dashed lines in Fig.~\ref{FIG:TEN}, show that the $W^\pm Z$
background is unlikely to be a problem in $W^+W^-$ production. For the cuts
chosen, it is at least one order of magnitude smaller than the $W^+W^-$
signal.

The top quark and $W^\pm Z$ backgrounds contribute to
$\ell_1^+\ell_2^-p\llap/_T+X$ production for all lepton flavor
combinations, $\ell_{1,2}=e,\,\mu$. Other background processes such as
$ZZ$ production where one of the $Z$ bosons decays into charged leptons,
$Z\to\ell^+\ell^-$, and the other into neutrinos, $Z\to\bar\nu\nu$,
contribute only for $\ell_1=\ell_2$. The transverse momentum
distribution of the charged lepton pair in $ZZ\to e^+e^-p\llap/_T+X$
is given by the dot dashed lines in Fig.~\ref{FIG:TEN}. The $p_T(e^+e^-)
$ distribution from $ZZ$ production is seen to be significantly harder
than that from $p\,p\hskip-7pt\hbox{$^{^{(\!-\!)}}$} \to W^+W^-$. For
$p_T(e^+e^-)$ values larger than about 120~GeV, the $ZZ$ background is
larger than the $W^+W^-$ signal, thus reducing the sensitivity to
anomalous $WWV$ couplings.

The production of $Z$ bosons accompanied by one or more jets also
contributes to the background in $\ell^+\ell^-p\llap/_T+X$ production,
if the rapidity of one of the jets is outside the range covered by the
detector and thus contributes to the missing transverse momentum. For a
realistic assessment of this background, a full-fledged Monte Carlo
simulation is required. Here, for a rough estimate, we use a simple
parton level calculation of $Z+1$~jet production.
For a jet, {\it i.e.} a quark or gluon, to be
misidentified as $p\llap/_T$ at the Tevatron (LHC), we require that the
jet pseudorapidity be $|\eta(j)|>3$~(4.5). The hadron calorimeters of CDF
and D\O\ cover the region up to $|\eta|\approx 4$~\cite{CAL}, and the
LHC experiments are designing their calorimeters to extend out to
$|\eta|\approx 5$~\cite{ATLAS,CMS}. Our results are thus expected to
be conservative. The
$p_T(e^+e^-)$ distribution for $Z+1~{\rm jet}\to e^+e^-p\llap/_T$ in
Fig.~\ref{FIG:TEN} is represented by the dotted line. It drops very
quickly for lepton pair transverse momenta above 30~GeV (50~GeV) at the
Tevatron (LHC) and does not affect the sensitivity to anomalous
couplings in any way.

Backgrounds where the $\ell^+\ell^-$ pair originates from a $Z$ boson
can be easily suppressed by requiring that
\begin{eqnarray}
|m(\ell^+\ell^-) -M_Z|>10~{\rm GeV}.
\label{EQ:MASSZ}
\end{eqnarray}
While this cut almost completely eliminates those
background processes, it hardly affects the $W^+W^-$ signal. This is
demonstrated in Fig.~\ref{FIG:ELEVEN}, where we compare the lowest order
lepton pair transverse momentum distribution with and without the cut on
the invariant mass of the lepton pair for $W^+W^-$ production in the SM.
The effect of the $m(e^+e^-)$ cut is particularly small at high
$p_T(e^+e^-)$ values, and therefore does not noticeably
influence the sensitivity
to anomalous $WWV$ couplings.

Numerous other processes contribute to the background in the $\ell_1^+
\ell_2^-p\llap/_T+X$ channels. In order not to overburden the figure,
the $p_T(e^+e^-)$ differential cross sections from these processes are
not included in Fig.~\ref{FIG:TEN}. The rate for associated production
of $W$ bosons
and top quarks, $p\,p\hskip-7pt\hbox{$^{^{(\!-\!)}}$} \rightarrow
W^-t+X,~W^+\bar t+X\to\ell_1^+\ell_2^-p\llap/_T+X$, is about a
factor~50 (100) smaller than the $t\bar t$ cross section at the Tevatron
(LHC)~\cite{CAVA,LAD} and therefore does not represent a problem.
Due to the relatively high lepton
and missing transverse momentum cuts we impose (see Sec.~IIIA), the
$Z+X\to\tau^+\tau^-+X\to e^+e^-p\llap/_T+X$ background is
substantially suppressed. Furthermore, the $p_T(e^+e^-)$ distribution
from $Z\to\tau^+\tau^-$ decays falls very steeply; for $p_T(e^+e^-)>
50$~GeV the $Z$ boson must either be far off-shell, or be accompanied by
a high $p_T$ jet. Using the ``poor man's shower'' approach~\cite{PMS} to
simulate the transverse motion of the $Z$ boson, we find that the
$Z+X\to\tau^+\tau^-+X\to e^+e^-p\llap/_T+X$ background to be at least a
factor~5 (10) smaller than the $W^+W^-$ signal at the Tevatron (LHC)
over the entire $p_T(e^+e^-)$ range. The background from $\bar bb$,
$\bar cc$, $Wg\to t\bar b$~\cite{YUAN,ALAN}, $q\bar q'\to
t\bar b$~\cite{ALAN,CORT}, $Wc$~\cite{CHARM}
or $W\bar bb$, $W\bar cc$ production is negligible (small)
at the Tevatron~\cite{TOPMASS1,TOPMASS2,MANG} (LHC~\cite{ATLAS,CAVA}) after
lepton isolation cuts are imposed.

In contrast to the charm and bottom background, the top quark background
is only insignificantly reduced by lepton isolation cuts. However, the
$b$-quarks produced in top quark decays frequently lead to one or
two hadronic jets~\cite{HTOP}, and a 0-jet requirement can be used to suppress
the $t\bar t$, as well as the $Wt+X$, rate. The decomposition of the
$p_T(e^+e^-)$ differential
cross section in $t\bar t$ production at lowest order into 0-jet, 1-jet,
and 2-jet exclusive cross sections at the Tevatron and LHC for
$m_t=176$~GeV is shown in Fig.~\ref{FIG:TWELVE}, using the jet definitions of
Eqs.~(\ref{EQ:TEVJET}) and~(\ref{EQ:LHCJET}) together with a jet
clustering algorithm.  The clustering algorithm merges the
$b$- and $\bar b$-quark into one jet if their separation is $\Delta
R(b,\bar b)<0.4$ and their combined transverse momentum is
larger than the jet-defining $p_T$ threshold.

At Tevatron energies, $t\bar t$ production predominantly leads to
$W^+W^-+2$~jet events. Less than 1\% of the events have no jet with $p_T(j)
>20$~GeV. At the LHC, for lepton pair transverse momenta smaller than
about 300~GeV, the fraction of $W^+W^-+2$~jet and $W^+W^-+1$~jet from
$t\bar t$ production is roughly equal.
At very large values of $p_T(e^+e^-)$, the
majority of all $t\bar t$ events contain two jets. Approximately 10\%
of all events have no jet with a transverse momentum in excess of
50~GeV.

As an alternative to a jet veto, a cut on the transverse momentum of the
hadrons, $p_T(h)$, can be imposed in order to suppress the top quark
background~\cite{DOWW}. The transverse momentum vector of the
hadrons is related to the other transverse momenta
in an $e^+e^-p\llap/_T+X$ event through the equation
\begin{eqnarray}
\hbox{\bf p}_T(h)=-\left[\hbox{\bf p}_T(e^+) + \hbox{\bf p}_T(e^-) +
\hbox{\bf p\llap/}_T\right].
\label{EQ:PTHAD}
\end{eqnarray}
In contrast to a jet veto requirement, a cut on $p_T(h)$ is independent
of the jet definition, in particular the jet cone size. It also
significantly reduces the dependence on the jet energy corrections. For
$t\bar t$ production in the dilepton channel, at LO, $p_T(h)=p_T(\bar
bb)$, the transverse momentum of the $\bar bb$ pair. For
$W^+W^-+X\to\ell_1^+\ell_2^-p\llap/_T+X$, at NLO, $p_T(h)$ coincides
with the jet transverse momentum. In this case, a jet veto and a cut
on $p_T(h)$ are equivalent.

The effect of a $p_T(h)<20$~GeV (50~GeV) cut at the Tevatron (LHC) is
shown by the long dashed lines in Fig.~\ref{FIG:TWELVE}. Clearly, at the
Tevatron, the $p_T(h)$ cut is considerably less efficient than a 0-jet
requirement with a cut on the jet $p_T$ equal to the cut imposed on
$p_T(h)$. At the LHC, the jet veto is only slightly more efficient than
a cut on the transverse momentum of the hadrons. Results which are
qualitatively very similar to those shown in Fig.~\ref{FIG:TWELVE} are
obtained for $m_t=200$~GeV. For the larger top quark mass, the $t\bar
t$ differential cross section is approximately a factor~1.3 to 1.5
smaller in the high $p_T(e^+e^-)$ tail, if a 0-jet requirement or a
$p_T(h)$ cut are imposed.

In Figs.~\ref{FIG:THIRTEEN} and~\ref{FIG:FOURTEEN}, we compare the
$p_T(e^+e^-)$ differential cross section of the $W^+W^-$ signal
with the residual $t\bar t$ background at Tevatron and LHC energies,
respectively, for two jet-defining $p_T$ thresholds. For the jet
definition of Eq.~(\ref{EQ:TEVJET}), a jet veto is seen to reduce the
$t\bar t$ background at the Tevatron to a few per cent of the signal (see
Fig.~\ref{FIG:THIRTEEN}a). On the other hand,
if a $p_T(h)<20$~GeV cut is imposed, the top
quark background is still about half as large as the $W^+W^-$ signal in
the high $p_T(e^+e^-)$ tail. For $p_T(h)<10$~GeV, the $t\bar t$ rate is
approximately one order of magnitude below the $W^+W^-$ signal cross
section. At the LHC (Fig.~\ref{FIG:FOURTEEN}), neither a cut on the
transverse momentum
of the hadrons of $p_T(h)<50$~GeV nor a jet veto with the same 50~GeV
$p_T$ threshold are sufficient to reduce the $t\bar t$ rate to below the
$W$ pair signal. If the threshold of the $p_T(h)$ or jet veto cut can be
lowered to 30~GeV, the top quark background can be reduced by an
additional factor~2 to~5. Nevertheless, the residual $t\bar t$ rate is
still larger than the $W^+W^-$ cross section for large values of
$p_T(e^+e^-)$.

It is difficult to further reduce the top quark background at the LHC.
Once a jet veto is imposed, the characteristics of $W^+W^-$ signal and
$t\bar t$ background events are very similar. To suppress the $t\bar t$
cross section to below the $W^+W^-$ rate, one would need to reduce the
transverse momentum threshold in the jet veto or the $p_T(h)$ cut to a
value considerably below 30~GeV. This is probably only feasible if the LHC
is operated significantly below its design luminosity of ${\cal
L}=10^{34}~{\rm cm}^{-2}~{\rm s}^{-1}$.

In our estimate of the top quark background, we have calculated the $t\bar
t$ cross section to lowest order in $\alpha_s$. Higher order QCD corrections
affect the $t\bar t$ differential cross sections only
slightly~\cite{NLOTOP} and therefore do not appreciably
change the results shown in Figs.~\ref{FIG:TEN} --~\ref{FIG:FOURTEEN}.

In Fig.~\ref{FIG:FIFTEEN}, finally, we display the $p_T(e^+e^-)$
distribution for $p\,p\hskip-7pt\hbox{$^{^{(\!-\!)}}$}
\to e^+e^-p\llap/_T+0$~jet where we have added the differential cross
sections of the $W^+W^-+0$~jet signal and the residual top quark
background. Results are displayed for the SM and for anomalous $WWV$
couplings in the HISZ scenario~\cite{HISZ} (see Sec.~IIB). So
far, in order to investigate how the differential cross sections depend
on the non-standard $WWV$ couplings,
we have assumed that only one anomalous coupling at a time
is non-vanishing. In a realistic model, there is no reason to expect
that this is the case. The scenario of Ref.~\cite{HISZ} provides an
example of a model in which both $WW\gamma$ and $WWZ$ anomalous couplings
are simultaneously non-zero, thus making it possible to study the
interference effects between the different non-standard couplings.
Furthermore, the number of independent $WWV$ couplings in this scenario
can be reduced from five to two [see Eqs.~(\ref{EQ:HISZ1})
--~(\ref{EQ:HISZ3})] by imposing one simple additional constraint.
The dashed and dotted lines in Fig.~\ref{FIG:FIFTEEN} display the
$p_T(e^+e^-)$ distribution of signal plus background for two sets of
non-standard couplings fulfilling Eqs.~(\ref{EQ:HISZ1})
--~(\ref{EQ:HISZ3}).
For simplicity, only one of the two independent couplings is
allowed to differ from its SM value at a time. The figure shows that
at the Tevatron the sensitivity to
anomalous $WWV$ couplings remains virtually unaffected by the $t\bar
t\to e^+e^-p\llap/_T+0$~jet background, whereas it is significantly
reduced at the LHC.

\subsection{$W\to\tau\nu$ Decay Modes}

So far, we have completely ignored the contributions from decay modes
where one or both charged leptons in the final state originate from
$W\to\tau\nu_\tau\to e\nu_e\bar\nu_\tau\nu_\tau$.
Experimentally, it is difficult to separate the $W\to\tau\nu$ and $W\to
e\nu$ channels if the $\tau$ decays into leptons only. It is
straightforward to implement $\tau$ decays into our calculation; one
simply replaces the $W\to e\nu$ decay current, $J_\mu(k)$, with the
$W\to\tau\nu_\tau\to e\nu_e\nu_\tau\nu_\tau$ decay current,
$D^\tau_\mu(k)$.

In Fig.~\ref{FIG:FIFTEENA}, we compare the LO $p_T(e^+e^-)$
spectrum of $e^+e^-$ pairs where one (dashed lines) or both leptons
(dotted lines) originate from $\tau$ decays with the distribution where
both leptons originate from ``prompt'' $W\to e\nu$ decays. Leptons
originating from $\tau$ decays are significantly softer than those from
prompt decays. The $W\to\tau\nu$ decay modes,
therefore, are significantly suppressed by the $p_T$ cut which we
impose on charged leptons (see Sec.~IIIB). If both $W$'s decay into
$\tau$-leptons, the combined branching ratio of the subsequent $\tau$
decay, $[B(\tau\to e\nu_e\nu_\tau)]^2\approx 0.032$, further reduces the
contribution of this channel. As a result, the $p_T(e^+e^-)$
differential cross section where both leptons originate from $\tau$
decays is approximately 3~orders of magnitude below that from
prompt $e^+e^-$ pairs. Since the decay leptons are emitted roughly
collinear with the direction of the parent $\tau$-lepton, the
delicate balance of the transverse momenta of the leptons in the $W\to
e\nu$ case is preserved if both $W$'s decay into $\tau$-leptons, and the
slope of the $p_T(e^+e^-)$ distributions from $W^+W^-\to e^+\nu_e
e^-\bar\nu_e$ and $W^+W^-\to\tau^+\nu_\tau\tau^-\bar\nu_\tau$ are similar.

However, this is not the case if only one of the two $W$ bosons decays
into $\tau\nu$. The charged lepton from the decaying $\tau$-lepton is
typically much softer than that originating from $W\to e\nu$, thus
spoiling the balance of transverse momenta.
The resulting $p_T(e^+e^-)$ distribution is somewhat
harder than that from $W^+W^-\to e^+\nu_e e^-\bar\nu_e$. While the rate
of the $\tau$ decay mode is smaller by approximately one order of
magnitude at low values of $p_T(e^+e^-)$, it is larger than the
$W^+W^-\to e^+\nu_e e^-\bar\nu_e$ cross section for $p_T(e^+e^-)>200$~GeV
(250~GeV) at the Tevatron (LHC). Decay modes where one of the $W$ bosons
decays into $\tau\nu$ thus change the shape of the $p_T(e^+e^-)$
distribution, although considerably less than NLO QCD corrections do.

The NLO 0-jet $e^+e^-$ transverse momentum distributions are very
similar to the LO differential cross sections shown in
Fig.~\ref{FIG:FIFTEENA}. At the inclusive NLO level, or in the case of
non-zero anomalous $WWV$ couplings, the correlation of the charged
lepton transverse momenta found in the SM LO $W^+W^-\to e^+\nu_e
e^-\bar\nu_e$ case is not present and the $p_T(e^+e^-)$ differential
cross section for $W^+W^-\to e^\pm\nu_e\tau^\mp\nu_\tau$ is about one
order of magnitude below that from $W^+W^-\to e^+\nu_e e^-\bar\nu_e$
over the entire transverse momentum range considered. Contributions from
channels where one $W$ boson decays into a $\tau$-lepton thus
slightly reduce the
overall sensitivity to anomalous couplings.

\subsection{Sensitivity Limits}

We now proceed and derive sensitivity limits for anomalous $WWV$
couplings from $W^+W^-+X\to\ell_1^+\ell_2^-p\llap/_T+X$, $\ell_{1,2}=e,
\,\mu$, at the Tevatron and LHC. For the Tevatron we
consider integrated luminosities of 1~fb$^{-1}$, as envisioned for the
Main Injector era, and 10~fb$^{-1}$ (TeV*) which could be achieved
through additional upgrades of the Tevatron accelerator
complex~\cite{GPJ}. In the case of the LHC we use $\int\!{\cal
L}dt=10$~fb$^{-1}$ and 100~fb$^{-1}$~\cite{LHC}. To extract
limits, we shall sum over electron and
muon final states. Interference effects between different $WWV$
couplings are fully incorporated in our analysis. We derive limits for
the cases where either the $WW\gamma$, or the $WWZ$ couplings, only
are allowed to  differ from their SM values,
as well as for the HISZ scenario described at
the end of Sec.~IIB. Varying the $WW\gamma$ or $WWZ$ couplings
separately makes it possible to
directly compare the sensitivity of $W^+W^-$ production to these
couplings with that of $W\gamma$ and $WZ$ production.
Furthermore, the bounds derived in these limiting cases
make it easy to perform a qualitative estimate of sensitivity limits for any
model where the $WWZ$ and $WW\gamma$ couplings are related.
The HISZ scenario serves as a simple example of such a model. In the form we
consider here, only two of the
couplings are independent; see Eqs.~(\ref{EQ:HISZ1}) --~(\ref{EQ:HISZ3}).

To derive 95\%~CL limits we use the $p_T(\ell_1^+\ell_2^-)$
distribution and perform a $\chi^2$ test~\cite{ZGAM}, assuming that no
deviations from the SM predictions are observed in the experiments
considered. As we have seen, the $\ell_1^+\ell_2^-$ transverse momentum
distribution in general yields the best sensitivity bounds in the Born
approximation. Furthermore, we impose the cuts summarized in Sec.~IIIB. For
simplicity, we do not exclude the region around the $Z$ mass peak in
$m(\ell_1^+\ell_2^-)$ for
$\ell_1=\ell_2$, which is necessary to eliminate the background from
$ZZ\to\ell^+\ell^-p\llap/_T$. As we have demonstrated in
Fig.~\ref{FIG:ELEVEN}, such a cut does not noticeably influence the high
$p_T(\ell_1^+\ell_2^-)$ region from which most of the sensitivity to
anomalous $WWV$ couplings originates. We also ignore any contributions
from decay modes where one or both $W$'s decay into a $\tau$-lepton.
These modes affect the sensitivity to non-standard $WWV$ couplings only
insignificantly (see Sec.~IIIF). Since most background processes
can be removed by standard requirements such as an isolated charged
lepton cut, we concentrate on the $t\bar t$ background. For the top quark
mass we assume $m_t=176$~GeV. At the
Tevatron with 1~fb$^{-1}$ (10~fb$^{-1}$) we use a jet-defining $p_T$
threshold of 10~GeV (20~GeV), whereas we take 30~GeV (50~GeV) at the LHC
for 10~fb$^{-1}$ (100~fb$^{-1}$). Unless
explicitly stated otherwise, a dipole form factor ($n=2$) with scale
$\Lambda_{FF}=1$~TeV is assumed. The $p_T(\ell_1^+\ell_2^-)$
distribution is split into a certain number of bins. The number of bins
and the bin width depend on
the center of mass energy and the integrated luminosity. In each bin the
Poisson statistics are approximated by a Gaussian distribution. In order
to achieve a sizable counting rate in each bin, all events above a
certain threshold are collected in a single bin.
This procedure guarantees that a high statistical significance cannot
arise from a single event at large transverse momentum, where the
SM predicts, say, only 0.01 events. In order to derive
realistic limits we allow for a normalization uncertainty of 50\% in the
SM cross section. By employing more powerful statistical
tools than the simple $\chi^2$ test we performed~\cite{GREG}, it may be
possible to improve the limits we obtain.

In Figs.~\ref{FIG:SIXTEEN} and~\ref{FIG:SEVENTEEN}, and in
Table~\ref{TABLE1} we display sensitivity limits for the case where only
the $WWZ$ couplings are allowed to deviate from their SM values. The
cross section in each bin is a bilinear function of the anomalous couplings
$\Delta\kappa^0_Z$, $\lambda^0_Z$, and $\Delta g_1^{Z0}$. Studying the
correlations in the $\Delta\kappa^0_Z$ --~$\lambda^0_Z$, the
$\Delta\kappa^0_Z$ --~$\Delta g_1^{Z0}$,
and the $\Delta g_1^{Z0}$ --~$\lambda^0_Z$ planes is therefore sufficient
to fully include all interference effects between the various
$WWZ$ couplings. Figure~\ref{FIG:SIXTEEN} (\ref{FIG:SEVENTEEN}) shows
95\% CL contours in the three planes for the Tevatron (LHC) with
1~fb$^{-1}$ (10~fb$^{-1}$). Without a jet veto, inclusive NLO
corrections and the top quark background together reduce the
sensitivity obtained
from the LO $W^+W^-$ cross section by about a factor~2 to~5. Imposing a
jet veto, the $t\bar t$ background and the large QCD corrections at high
$\ell_1^+\ell_2^-$ transverse momenta are essentially eliminated at the
Tevatron, and the resulting limits are very similar to those obtained
from the LO analysis. At the LHC, the remaining top quark background still
has a non-negligible impact, reducing the limits obtained from the
analysis of $W^+W^-$ production at LO by a factor~1.5 --~2. The bounds
extracted from the LO $W^+W^-$ cross section represent the results for the
ideal case where all background can be completely removed. The limits
obtained without reducing the $t\bar t$ background and the NLO QCD
corrections, on the other hand, correspond to a `worst case scenario',
{\it i.e.}, the minimal sensitivity to
anomalous couplings which one should be able to reach.

More detailed information on how QCD corrections and the top quark
background influence
the limits which can be achieved on $WWZ$ couplings is provided in
Table~\ref{TABLE1}. At Tevatron energies, NLO QCD corrections reduce the
sensitivity by 5~--~10\%, while for the LHC the bounds obtained from the
inclusive NLO $W^+W^-$ cross section are typically a factor~2 worse than
those extracted using the LO cross section. A 10\% (factor~2) variation
in the 95\% CL limits is roughly equivalent to a factor 1.5 (16) in
integrated luminosity needed to compensate for the effect of the NLO
corrections. The limits found by imposing
a $p_T(h)$ cut and a jet veto requirement are almost identical at the
Tevatron. For LHC energies, the $p_T(h)$ cut yields bounds which are
20~--~40\% weaker than those extracted from the exclusive NLO $W^+W^-$
rate.

Terms in the amplitudes proportional to $\Delta g_1^Z$ grow like
$\sqrt{\hat s}/M_W$ while terms multiplying $\Delta\kappa_V$ and
$\lambda_V$ increase with $\hat s/M_W^2$. As a result, the limits which
can be achieved for $\Delta g_1^Z$ are significantly weaker than the
bounds obtained for $\Delta\kappa_Z$ and $\lambda_Z$. Our limits also fully
reflect the sign-dependence of the differential cross sections for
$\Delta g_1^Z$ and $\Delta\kappa_V$ noted earlier.

Limits for the cases in which the $WW\gamma$ couplings are varied (assuming
SM $WWZ$ couplings) and the HISZ scenario are shown in
Figs.~\ref{FIG:EIGHTEEN}
and~\ref{FIG:NINETEEN}, and Tables~\ref{TABLE2} and~\ref{TABLE3}. We
only display the limits for the NLO 0-jet case, including the residual
$t\bar t$ background, in these figures and tables. In
Fig.~\ref{FIG:EIGHTEEN} we compare the limits for the three different
cases for a fixed integrated luminosity. Due to
the smaller overall $WW\gamma$ and photon fermion couplings, the bounds
on $\Delta\kappa_\gamma$ and $\lambda_\gamma$ are about a factor~1.5
to~3 weaker than the limits obtained for $WWZ$ couplings. As a result of
the assumed relations between the $WW\gamma$ and $WWZ$ couplings [see
Eqs.~(\ref{EQ:HISZ1}) --~(\ref{EQ:HISZ3})], we find limits on
$\lambda_\gamma$ ($\Delta\kappa_\gamma$) in the HISZ scenario, which
are somewhat better (worse) than those obtained for $\lambda_Z$
($\Delta\kappa_Z$) when only the $WWZ$ couplings are varied.
The CDF and D\O\ Collaborations have derived 95\% CL limit contours for the
$WWV$ couplings from $W^+W^-$ production~\cite{CDFWW,DOWW} for the case
$\Delta\kappa_Z=\Delta\kappa_\gamma$, $\lambda_Z=\lambda_\gamma$, and
$\Delta g_1^Z=0$. In this scenario, we find limits which are about
20~--~40\% better than those obtained for the case where only
$\Delta\kappa_Z$ and $\lambda_Z$ are allowed to deviate from their SM
values.

In Fig.~\ref{FIG:NINETEEN} we compare the bounds which can be achieved for
the HISZ scenario for different integrated luminosities and form factor
scales. Increasing
the integrated luminosity by one order of magnitude improves the
sensitivity limits by a factor~2.0 --~2.7 at the
Tevatron, and up to a factor of~1.8 at the LHC for the form factor scale
chosen. Due to the significantly higher residual top quark background,
the sensitivity limits which can be achieved at the LHC with
10~fb$^{-1}$ are only up to a factor~2 better than those found at the
Tevatron for the same integrated luminosity and form factor scale.

At Tevatron energies, the sensitivities achievable are insensitive to
the exact form and scale of the form factor for $\Lambda_{FF}>400$~GeV.
At the LHC, the situation is somewhat different and the sensitivity
bounds depend on the value chosen for $\Lambda_{FF}$. This is
illustrated in Fig.~\ref{FIG:NINETEEN}b and
Table~\ref{TABLE3}, where we display the limits which can
be achieved at the LHC with $\int\!{\cal L}dt=100$~fb$^{-1}$ and a form
factor scale of $\Lambda_{FF}=3$~TeV. The limits for the higher scale
are a factor~2.8 to~5 better than those found for $\Lambda_{FF}=1$~TeV
with the same integrated luminosity. For $\Lambda_{FF}>3$~TeV, the
sensitivity bounds depend only marginally on the form factor
scale~\cite{DPF}, due
to the very rapidly falling cross section at the LHC for parton center
of mass energies in the multi-TeV region. The dependence of the limits
on the cutoff scale $\Lambda_{FF}$ in the form
factor can be understood easily from Fig.~\ref{FIG:EIGHT}. The improvement in
sensitivity with increasing $\Lambda_{FF}$ is due to the additional events
at large $p_T(\ell_1^+\ell_2^-)$ which are suppressed by the
form factor if the scale $\Lambda_{FF}$ has a smaller value.

To a lesser degree, the bounds also depend on the power $n$ in the form
factor,
which we have assumed to be $n=2$. For example, the less drastic cutoff
for $n=1$ instead of $n=2$ in the form factor allows for additional
high $p_T(\ell_1^+\ell_2^-)$ events and therefore leads
to a slightly increased sensitivity to the low energy values of the
anomalous $WWV$ couplings. The sensitivity bounds listed in
Tables~\ref{TABLE1} --~\ref{TABLE3} can thus be taken as representative
for a wide class of
form factors, including the case where constant anomalous couplings are
assumed for $M_{WW}<\Lambda_{FF}$, but invariant masses above
$\Lambda_{FF}$ are ignored in deriving the sensitivity
bounds~\cite{FLS}.

{}From our studies we conclude that at the TeV* the $WWV$ couplings can be
probed with an accuracy of 10~--~60\%, except for $\Delta g_1^Z$. At the
LHC, with $\int\!{\cal L}dt=100$~fb$^{-1}$, $\Delta\kappa_V^0$ and
$\lambda_V^0$ can be determined with an uncertainty of a few per cent,
whereas $\Delta g_1^{Z0}$ can be measured to approximately 0.2, with
details depending on the form factor scale assumed.
For a top quark mass of $m_t=200$~GeV, we find sensitivity bounds which
are slightly better than those shown in Figs.~\ref{FIG:SIXTEEN}
--~\ref{FIG:NINETEEN} and Tables~\ref{TABLE1} --~\ref{TABLE3}.
Limits derived from the transverse momentum distribution of the
individual charged leptons are weaker by approximately a factor~1.5 than
those extracted from the $p_T(\ell_1^+\ell_2^-)$ spectrum. We have not
studied the sensitivities which can be achieved in the current Tevatron
collider run in detail. For an integrated luminosity of about 100~pb$^{-1}$
the limits which one can hope to achieve are approximately a factor two
to three worse than those found for 1~fb$^{-1}$.

The results shown in Figs.~\ref{FIG:SIXTEEN} --~\ref{FIG:NINETEEN} and
Tables~\ref{TABLE1} --~\ref{TABLE3} should be compared with the
sensitivities expected in other channels~\cite{DPF,ATLAS,BHO1}, and in $W$
pair production at LEP~II~\cite{DPF,BIL,Sek}, and a linear $e^+e^-$
collider~\cite{NLCWW}. The limits which we obtain for the $WW\gamma$
couplings at the Tevatron, assuming a SM $WWZ$ vertex function, are a
factor~1.7 --~4.4 weaker than those projected from $W^\pm\gamma$ production
with $W\to e\nu$~\cite{DPF}, mostly due to the smaller event rate. At
the LHC, with 100~fb$^{-1}$ and $\Lambda_{FF}=3$~TeV, the limits on
$\Delta\kappa^0_\gamma$ ($\lambda^0_\gamma$) are a factor~1.5 to~2
($\sim 3$) better (worse) than those expected from $W\gamma$
production~\cite{DPF,ATLAS}.
The higher sensitivity of $W$ pair production to $\Delta\kappa_\gamma$
can be traced to the high energy behaviour of the terms proportional to
$\Delta\kappa_V$ in the helicity amplitudes. As mentioned in the
Introduction, these terms increase proportional to $\hat s/M_W^2$ in
$W^+W^-$ production, whereas they grow only like $\sqrt{\hat s}/M_W$ in
$p\,p\hskip-7pt\hbox{$^{^{(\!-\!)}}$} \rightarrow W^\pm\gamma,\, W^\pm
Z$.

The bounds we obtain for the $WWZ$ couplings, assuming a SM $WW\gamma$
vertex, can be compared directly with the sensitivity limits calculated
for $W^\pm Z\to\ell_1^\pm\nu_1\ell_2^+\ell_2^-$ in Ref.~\cite{BHO1}.
The bounds for
$\lambda_Z$ from $W^+W^-$ and $W^\pm Z$ production are very similar. At
the LHC, the larger cross section for $W^+W^-$ production is compensated
by the considerable top quark background which remains even after a jet
veto has been imposed. For Tevatron (LHC) energies, the sensitivity limits
for $\Delta\kappa_Z$ from $W$ pair production are approximately a
factor~3 (2~--~7) better than those which can be achieved in $p\bar p\to
WZ$ ($pp\to WZ$),
whereas the bounds for $\Delta g_1^Z$ from $WZ$ production are 3~--~4
(7~--~34) times more stringent than those extracted from the $W^+W^-$
channel for the parameters chosen. $WW$ and $WZ$ production at hadron
colliders thus yield complementary information on $\Delta g_1^Z$ and
$\Delta\kappa_Z$. The limits fully reflect the high energy behaviour of
the individual helicity amplitudes for the two processes. Terms
proportional to $\lambda_Z$ increase in both cases like $\hat s/M_W^2$.
On the other hand, the leading $\Delta g_1^Z$ ($\Delta\kappa_Z$) terms
in $WZ$ ($WW$) production
grow faster with energy [$\sim\hat s/M_W^2$] than in $WW$ ($WZ$) production
[$\sim\sqrt{\hat s}/M_W$].

In the HISZ scenario, $WW$ production leads
to bounds for $\Delta\kappa_\gamma$ which, at the Tevatron (LHC), are
up to factor of two (five)
weaker than those obtained in $W\gamma$ and $WZ$ production~\cite{DPF}.
The limits on $\lambda_\gamma$ from $W$ pair production at the Tevatron
(LHC) in this model are slightly better (worse) than those derived from
$W^\pm Z\to\ell_1^\pm\nu_1\ell_2^+\ell_2^-$.

As has been demonstrated by the CDF Collaboration~\cite{CDFWW}, useful
limits on the $WWV$ couplings can also be derived from $WW,\,
WZ\to\ell\nu jj$ and $WZ\to\ell^+\ell^- jj$ at large di-jet transverse
momenta, $p_T(jj)$. Decay modes where one of the vector bosons
decays hadronically have a considerably larger branching ratio
than the $W^+W^-\to\ell_1^+\nu_1\ell_2^-\bar\nu_2$ channel and thus yield
higher rates. On the other hand, a jet veto cannot be utilized to reduce
the top background for the semihadronic final states. Due to the very
large $t\bar t$ background at the LHC, decay modes where one of the
vector bosons decays into hadrons are therefore only useful at Tevatron
energies where the total $t\bar t$ and $W^+W^-$ production rates are
comparable. Here, a sufficiently large $p_T(jj)$ cut eliminates the QCD
$W/Z+$~jets background and the SM signal, but retains good sensitivity
to anomalous $WWV$ couplings. The value of the $p_T(jj)$ cut varies with
the integrated luminosity assumed. Simulations of the sensitivities which
may be expected in the HISZ scenario for $WW,\,WZ\to\ell\nu jj$ and
$WZ\to\ell^+\ell^- jj$
in future Tevatron experiments show~\cite{DPF} that, for 1~fb$^{-1}$,
the semihadronic final states yield bounds for $\Delta\kappa_\gamma$
which are roughly a factor two more stringent as those from
$W^+W^-\to\ell_1^+\nu_1\ell_2^-\bar\nu_2$, whereas the limits on
$\lambda_\gamma$ are very similar. With growing integrated luminosity,
it is necessary to raise the $p_T(jj)$ cut to eliminate the $W/Z+$~jets
background. For increasing values of $p_T(jj)$, more and more jets
tend to coalesce. At $\int\!{\cal L}dt\ge 10$~fb$^{-1}$, jet
coalescing severely degrades the limits on anomalous $WWV$ couplings
which can be achieved. With growing integrated luminosity, $W^+W^-$
production in the all leptonic channels thus becomes increasingly potent
in constraining the $WWV$ vertices.

The sensitivities in the HISZ scenario which one hopes to achieve
from $p\bar p\to W^+W^-+0~{\rm jet}\to\ell_1^+\ell_2^-p\llap/_T+0$~jet
(short dashed line) and the other di-boson production channels (adopted
from Ref.~\cite{DPF}) at
the Tevatron with 10~fb$^{-1}$ are summarized in Fig.~\ref{FIG:TWENTY}
and compared with the expectations from $e^+e^-\to W^+W^-\to\ell\nu jj$
at LEP~II for $\sqrt{s}=190$~GeV and $\int\!{\cal L}dt=500$~pb$^{-1}$
(long dashed line)~\cite{LEPII}. A similar comparison, with very similar
conclusions, can be carried out for the more conservative choices of an
integrated luminosity of 1~fb$^{-1}$ at the Tevatron, and a center of
mass energy of $\sqrt{s}=176$~GeV at LEP~II~\cite{LEPII}.
While $W\gamma$ production is seen to
yield the best bounds at the Tevatron over a large fraction of the
parameter space, it is clear that the limits obtained from the
various processes are all of similar magnitude. In particular, the
limits from the all leptonic decays of $W$ pairs are seen to be
comparable to those from the other $WW$ and $WZ$ channels for a
significant part of the $\Delta\kappa_\gamma^0$ -- $\lambda_\gamma^0$
plane. Performing a global analysis of all di-boson production channels
thus is expected to result in a significant improvement of the
sensitivity bounds which can be achieved.

Figure~\ref{FIG:TWENTY} also demonstrates that the limits from di-boson
production at the Tevatron and $W^+W^-$ production at LEP~II are quite
complementary. The contour for $e^+e^-\to W^+W^-\to\ell\nu jj$ in
Fig.~\ref{FIG:TWENTY} has been adopted from Ref.~\cite{DPF}, and is
based on an analysis which takes into account initial state radiation
and finite detector resolution effects, together with ambiguities in
reconstructing the $W$ decay angles in hadronic $W$ decays in absence of
a readily recognizable quark tag. Information on the $WWV$ couplings in
$e^+e^-\to W^+W^-\to\ell^\pm\nu jj$ is extracted from the angular
distribution of the final state fermions. Of the three final states
available in $W$ pair production,
$\ell_1\nu_1\ell_2\nu_2$, $\ell\nu jj$, $\ell=e,\,\mu$, and $jjjj$, the
$\ell\nu jj$ channel yields the best sensitivity bounds. The purely
leptonic channel is plagued
by a small branching ratio ($\approx 4.7\%$) and by reconstruction
problems due to the presence of two neutrinos. In the $jjjj$ final state
it is difficult to discriminate the $W^+$ and $W^-$ decay products. Due
to the resulting ambiguities in the $W^\pm$ production and decay angles,
the sensitivity bounds which can be achieved from the 4-jet final state
are a factor 1.5 --~2 weaker than those found from analyzing the
$\ell\nu jj$ state~\cite{Sek}.

At the
NLC, the $WWV$ couplings can be tested with a precision of
$10^{-3}$ or better. Details depend quite sensitively on the center of
mass energy and the integrated luminosity of the NLC~\cite{NLCWW}.

\section{SUMMARY}

$W^+ W^-$ production in hadronic collisions provides an opportunity to
probe the structure of the $WW\gamma$ and $WWZ$ vertices in a direct and
essentially model independent way. In contrast to other di-boson
production processes at hadron or $e^+e^-$ colliders, the reaction
$p\,p\hskip-7pt\hbox{$^{^{(\!-\!)}}$} \rightarrow W^+ W^- \rightarrow
\ell_1^+ \nu_1 \ell_2^- \bar \nu_2$ offers the possibility to {\sl
simultaneously} probe the high energy
behaviour and, at least indirectly, the helicity structure of the
$W^+W^-$ production amplitudes using the same observable. Usually,
information on the high energy behaviour of the di-boson production
amplitudes is obtained from transverse momentum and invariant mass
spectra, whereas angular distributions are used to probe the helicity
structure~\cite{LAGRANGIAN}.

Previous studies of $p\,p\hskip-7pt\hbox{$^{^{(\!-\!)
}}$} \rightarrow W^+ W^-$~\cite{FIRSTWW,BS,WWAC,HWZ,EHLQ} have been
based on leading order calculations. In this paper we have presented
an ${\cal O}(\alpha_s)$ calculation of the reaction
$p\,p\hskip-7pt\hbox{$^{^{(\!-\!)}}$} \rightarrow W^+ W^- + X
\rightarrow \ell_1^+ \nu_1 \ell_2^- \bar \nu_2 + X$ for general,
$C$ and $P$ conserving, $WW\gamma$ and
$WWZ$ couplings, using a combination of analytic and Monte Carlo
integration techniques. The leptonic decays $W \rightarrow \ell \nu$
have been included in the narrow width approximation in our calculation.
Decay spin correlations are correctly taken into account in the calculation,
except in the finite virtual contribution. The finite virtual
correction term contributes only at the few per cent level to the total
NLO cross section, thus decay spin correlations can be safely ignored here.
The calculation presented here complements earlier ${\cal O}(\alpha_s)$
calculations of $W^\pm\gamma$~\cite{BHO} and $W^\pm Z$~\cite{BHO1}
production at hadron colliders for general $C$ and $P$ conserving
anomalous $WWV$ couplings ($V=\gamma,\,Z$).

In the past, the all leptonic $W^+W^-$ decay channels have not been
considered in detail, due to the large $t\bar t$ background and event
reconstruction problems. The presence of two neutrinos in the event
makes it impossible to reconstruct the $WW$ invariant mass or the $W$
transverse momentum distribution. We have found that the limited
information available for the final state
does not reduce the sensitivity to anomalous couplings seriously when
the transverse momentum distribution of the charged lepton pair, or
equivalently, the missing $p_T$ distribution are considered. In
contrast to other distributions, the lepton pair transverse momentum
$p_T(\ell_1^+\ell_2^-)$ distribution is not only sensitive to the high
energy behaviour of the $W^+W^-$ production amplitudes, but
also provides indirect information on
the helicities of the $W$ bosons, which are strongly correlated in $W$
pair production in the SM~\cite{LAGRANGIAN,BS,HEL} (see Sec.~IIIC). The
correlation of the weak
boson helicities, together with the $V-A$ structure of the $We\nu$
coupling and the $2\to 2$ kinematics of leading order $W$ pair
production, causes a tendency for the
transverse momentum vectors of the two charged leptons to cancel, with a
corresponding sharp drop in the leading order SM $p_T(\ell_1^+\ell_2^-)$
distribution at high transverse momenta. Anomalous $WWV$ couplings do
not only change the high energy
behaviour of the helicity amplitudes, but also modify the correlation of
the $W$ helicities. As a result, the $p_T(\ell_1^+\ell_2^-)$
distribution, at leading order, exhibits a particularly pronounced
sensitivity to non-standard $WWV$ couplings. Decay channels where one of
the final state charged leptons originates from
$W\to\tau\nu_\tau\to\ell\nu_\ell\nu_\tau\bar\nu_\tau$, slightly modify
the shape of the $p_T(\ell_1^+\ell_2^-)$ distribution (see Sec.~IIIF).

The real emission processes, $q\bar q\to W^+W^-g$ and $qg\to W^+W^-q$,
which contribute to the ${\cal O}(\alpha_s)$ QCD corrections in $W$ pair
production, spoil the delicate balance of the charged lepton transverse
momenta. As a result, inclusive NLO QCD corrections to the
$p_T(\ell_1^+\ell_2^-)$ and $p\llap/_T$ distributions are very large
and may drastically reduce the sensitivity to non-standard $WWV$
couplings.  By imposing a jet veto, {\it i.e.}, by considering the exclusive
$W^+W^-+0$~jet channel instead of inclusive $W^+W^-+X$ production, the
QCD corrections are reduced to approximately 20\% of the LO cross
section, and the sensitivity to non-standard $WWV$ couplings is largely
restored. Furthermore, the
dependence of the NLO $W^+W^-+0$~jet cross section on the factorization
scale $Q^2$ is significantly reduced compared to that of the inclusive
NLO $W^+W^-+X$ cross section. Uncertainties which
originate from the variation of $Q^2$ will thus be smaller for
sensitivity bounds obtained from the $W^+W^-+0$~jet channel than for those
derived from the inclusive NLO $W^+W^-+X$ cross section.

A jet veto, or a cut on the hadronic transverse momentum, $p_T(h)$, also
helps to control the $t\bar t$ background. Without imposing such a cut,
the top quark background is much larger than the $W^+W^-$ signal at high
$\ell_1^+\ell_2^-$ transverse momenta and one looses a factor~2 --~5
in sensitivity. The jet veto in general is more efficient than a
$p_T(h)$ cut in reducing the top quark background (see
Fig.~\ref{FIG:THIRTEEN}). In practice, this difference is not very
important. For realistic $p_T(j)$ and
$p_T(h)$ thresholds, the $t\bar t$ background can be almost
completely eliminated at Tevatron energies. At the LHC, for both methods
only a signal to
background ratio of ${\cal O}(1)$ can be achieved. The residual $t\bar t$
background weakens the sensitivity bounds on anomalous couplings by
about a factor~1.5 --~2. Overall, the improvement of the sensitivity
bounds resulting from a jet veto or a cut on the hadronic
transverse momentum is equivalent to roughly a factor~10 --~40 increase
in integrated luminosity.

Excluding the region around the $Z$ mass in
$m(\ell_1^+\ell_2^-)$ for $\ell_1=\ell_2$ eliminates the
$ZZ\to\ell^+\ell^-p\llap/_T$ background which otherwise
dominates over the $W^+W^-$ signal at large values of
$p_T(\ell_1^+\ell_2^-)$. This cut has almost no effect on the
high $\ell_1^+\ell_2^-$ transverse momentum tail.

Due to the larger coupling of the $Z$ boson to quarks and $W$ bosons,
$W^+W^-$ production is more sensitive to $WWZ$ couplings than $WW\gamma$
couplings. Terms proportional to $\Delta\kappa_V$ in the amplitude
grow like $\hat s/M_W^2$, where $\hat s$ is the parton center of mass
energy squared, whereas these terms only grow like $\sqrt{\hat s}/M_W$
in $W^\pm\gamma$ and $W^\pm Z$ production. $W^+W^-$ production therefore
is considerably more sensitive to $\Delta\kappa_V$ than
$p\,p\hskip-7pt\hbox{$^{^{(\!-\!)}}$} \to W^\pm\gamma,\,W^\pm Z$. For
example, at the Tevatron (LHC) with $\int\!{\cal L}dt=10$~fb$^{-1}$
(100~fb$^{-1}$), varying only the $WWZ$ couplings, $\Delta\kappa_Z^0$
can be measured with 20~--~30\% (up to 2~--~3\%) accuracy [95\% CL] in
$W$ pair production in the purely leptonic channels.
These bounds are a factor~2 --~7 better than those which can be achieved
in $WZ$ production. Similarly, $W$ pair production yields better limits
for $\Delta\kappa_\gamma$ than $W^\pm\gamma$ production at the LHC for a
form factor scale $\Lambda_{FF}>2$~TeV, if the $WW\gamma$ couplings only
are varied. The sensitivity bounds which can be achieved for
$\Delta\kappa_V$ at the LHC approach the level where one would hope to
see deviations from the SM if new physics with a scale of ${\cal
O}(1$~TeV) exists. $\lambda_V$ can be determined with an accuracy of
10~--25\% (0.9~--~9\%) at the Tevatron (LHC), whereas $\Delta g_1^Z$ can
be probed at best at the 50\% (20\%) level. At the LHC, the limits
depend significantly on the form factor scale assumed. Detailed results
are shown in Figs.~\ref{FIG:SIXTEEN} --~\ref{FIG:NINETEEN} and
Tables~\ref{TABLE1} --~\ref{TABLE3}.

In the HISZ scenario [see Eqs.~(\ref{EQ:HISZ1}) --~(\ref{EQ:HISZ3})],
$W$ pair production at the Tevatron and LEP~II yield 95\% CL limit
contours which are quite complementary (see Fig.~\ref{FIG:TWENTY}).

%
\acknowledgements

We would like to thank T.~Diehl, T.~Fuess, S.~Keller, Y.K.~Kim, E.~Laenen,
C.~Wendt, T.~Yasuda, C.P.~Yuan, and D.~Zeppenfeld for stimulating
discussions. One of us (U.B.) is grateful to the Fermilab Theory Group,
where part of this work was carried out, for its generous hospitality.
This work has been supported in part by Department of Energy
grant No.~DE-FG03-91ER40674. U.B. is supported in part by a SUNY Term Faculty
Development Award. T.H. is supported in part by a UC-Davis Faculty
Research Grant.

%
%

%
\newpage
%
\widetext
\newcommand{\crc}{\crcr\noalign{\vskip -8pt}}
\begin{table}
\caption{Sensitivities achievable at the 95\% confidence
level (CL) for the anomalous $WWZ$ couplings $\Delta g_1^{Z0}$,
$\Delta\kappa^0_Z$, and
$\lambda^0_Z$ a) in $\protect{p\bar p\rightarrow W^+W^- + X\rightarrow
\ell_1^+ \ell_2^-p\llap/_T +X}$, $\ell_{1,2}=e,\,\mu$, at the Tevatron
($\protect{\sqrt{s}=1.8}$~TeV) with $\int\!{\cal L}dt=1$~fb$^{-1}$, and
b) in $\protect{pp\rightarrow
W^+W^- + X\rightarrow \ell_1^+ \ell_2^-p\llap/_T +X}$ at the LHC
($\protect{\sqrt{s}=14}$~TeV) with $\int\!{\cal L}dt=10$~fb$^{-1}$.
The limits for
each coupling apply for arbitrary values of the two other couplings.
The $WW\gamma$ couplings are assumed to take their SM values.
For the form factor we use the form of Eq.~(\protect{\ref{EQ:AFORM}})
with $n=2$ and $\Lambda_{FF}=1$~TeV. The transverse momentum threshold
for the jet veto and the $p_T(h)$ cut is taken to be 10~GeV at the Tevatron,
and 30~GeV at the LHC. The $t\bar t$ cross section is calculated at LO
with $m_t=176$~GeV.
The cuts summarized in Sec.~IIIB are imposed. \protect{\\} }
\label{TABLE1}
\begin{tabular}{cccccc}
 & & & & &\\ [-7.mm]
\multicolumn{6}{c}{a) Tevatron, $\int\!{\cal L}dt=1$~fb$^{-1}$}\\
\multicolumn{1}{c}{}
&\multicolumn{1}{c}{$W^+W^-$}
&\multicolumn{1}{c}{$W^+W^-$}
&\multicolumn{1}{c}{$W^+W^-+t\bar t$}
&\multicolumn{1}{c}{$W^+W^-+t\bar t$}
&\multicolumn{1}{c}{$W^+W^-+t\bar t$}\\
\multicolumn{1}{c}{$WWZ$ coupling}
&\multicolumn{1}{c}{LO}
&\multicolumn{1}{c}{NLO incl.}
&\multicolumn{1}{c}{NLO incl.}
&\multicolumn{1}{c}{NLO 0-jet}
&\multicolumn{1}{c}{NLO $p_T(h)$ cut}\\
\tableline
 & & & & &\\ [-6.mm]
$\Delta g_1^{Z0}$ & $\matrix{+1.96 \crc -1.22}$ & $\matrix{+2.10 \crc -1.
38}$ & $\matrix{+3.19 \crc -2.73}$ & $\matrix{+2.05 \crc -1.31}$ &
$\matrix{+2.08 \crc -1.34}$ \\ [5.mm]
$\Delta\kappa^0_Z$ & $\matrix{+0.61 \crc -0.48}$ & $\matrix{+0.66
\crc -0.51}$ & $\matrix{+1.22 \crc -0.99}$ & $\matrix{+0.66 \crc -0.
51}$ & $\matrix{+0.66 \crc -0.52}$ \\ [5.mm]
$\lambda^0_Z$ & $\matrix{+0.29 \crc -0.35}$ & $\matrix{+0.32
\crc -0.37}$ & $\matrix{+0.61 \crc -0.70}$ & $\matrix{+0.32 \crc -0.36}$
& $\matrix{+0.32 \crc -0.37}$ \\ [5.mm]
\tableline
\tableline
 & & & & &\\ [-7.mm]
\multicolumn{6}{c}{b) LHC, $\int\!{\cal L}dt=10$~fb$^{-1}$}\\
\multicolumn{1}{c}{}
&\multicolumn{1}{c}{$W^+W^-$}
&\multicolumn{1}{c}{$W^+W^-$}
&\multicolumn{1}{c}{$W^+W^-+t\bar t$}
&\multicolumn{1}{c}{$W^+W^-+t\bar t$}
&\multicolumn{1}{c}{$W^+W^-+t\bar t$}\\
\multicolumn{1}{c}{$WWZ$ coupling}
&\multicolumn{1}{c}{LO}
&\multicolumn{1}{c}{NLO incl.}
&\multicolumn{1}{c}{NLO incl.}
&\multicolumn{1}{c}{NLO 0-jet}
&\multicolumn{1}{c}{NLO $p_T(h)$ cut}\\
\tableline
 & & & & &\\ [-6.mm]
$\Delta g_1^{Z0}$ & $\matrix{+0.55 \crc -0.27}$ & $\matrix{+0.56 \crc -0.
61}$ & $\matrix{+1.19 \crc -1.57}$ & $\matrix{+0.81 \crc -0.50}$ &
$\matrix{+0.95 \crc -0.68}$ \\ [5.mm]
$\Delta\kappa^0_Z$ & $\matrix{+0.129 \crc -0.067}$ & $\matrix{+0.207
\crc -0.129}$ & $\matrix{+0.364 \crc -0.291}$ & $\matrix{+0.187 \crc -0.
123}$ & $\matrix{+0.217 \crc -0.156}$ \\ [5.mm]
$\lambda^0_Z$ & $\matrix{+0.043 \crc -0.045}$ & $\matrix{+0.078
\crc -0.090}$ & $\matrix{+0.138 \crc -0.146}$ & $\matrix{+0.063 \crc -0.
071}$ & $\matrix{+0.076 \crc -0.084}$ \\ [5.mm]
\end{tabular}
\end{table}
\newpage
\begin{table}
\caption{Sensitivities achievable at the 95\% confidence
level (CL) for anomalous $WWV$ couplings ($V=\gamma,\, Z$) in $\protect{p\bar
p\rightarrow W^+W^- + 0~{\rm jet}\rightarrow \ell_1^+ \ell_2^-
p\llap/_T+0}$~jet, $\ell_{1,2}=e,\,\mu$, at NLO for the Tevatron
a) for $\int\!{\cal L}dt=1$~fb$^{-1}$ and b) for $\int\!{\cal
L}dt=10$~fb$^{-1}$, including the residual background from $t\bar t$
production. Limits are shown for the case where only the
$WW\gamma$ or $WWZ$ couplings are allowed deviate from their SM values,
and for the HISZ scenario where we assume $\Delta\kappa_\gamma$ and
$\lambda_\gamma$ as the independent couplings [see
Eqs.~(\protect{\ref{EQ:HISZ1}}) --~(\protect{\ref{EQ:HISZ3}})].
Interference effects between those couplings which are varied are fully
taken into account.
For the form factors we use the form of Eq.~(\protect{\ref{EQ:AFORM}})
with $n=2$ and $\Lambda_{FF}=1$~TeV. The transverse momentum threshold
for the jet veto and the $p_T(h)$ cut is taken to be 10~GeV for
$\int\!{\cal L}dt=1$~fb$^{-1}$, and 20~GeV for 10~fb$^{-1}$. The $t\bar
t$ cross section is calculated at LO with $m_t=176$~GeV.
The cuts summarized in Sec.~IIIB are imposed. \protect{\\ [3.mm]} }
\label{TABLE2}
\begin{tabular}{cccc}
& & & \\[-6.mm]
\multicolumn{4}{c}{a) Tevatron, $\int\!{\cal L}dt=1$~fb$^{-1}$}\\
\multicolumn{1}{c}{coupling}
&\multicolumn{1}{c}{$WW\gamma$}
&\multicolumn{1}{c}{$WWZ$}
&\multicolumn{1}{c}{HISZ scenario}\\
\tableline
 & & & \\ [-6.mm]
$\Delta g_1^{Z0}$ & -- & $\matrix{+2.05 \crc -1.31}$ & -- \\ [5.mm]
$\Delta\kappa^0_V$ & $\matrix{+1.30 \crc -0.92}$ & $\matrix{+0.66
\crc -0.51}$ & $\matrix{+0.85 \crc -0.51}$ \\ [5.mm]
$\lambda^0_V$ & $\matrix{+0.58 \crc -0.51}$ & $\matrix{+0.32
\crc -0.36}$ & $\matrix{+0.22 \crc -0.20}$ \\ [5.mm]
\tableline
\tableline
& & & \\[-6.mm]
\multicolumn{4}{c}{b) Tevatron, $\int\!{\cal L}dt=10$~fb$^{-1}$}\\
\multicolumn{1}{c}{coupling}
&\multicolumn{1}{c}{$WW\gamma$}
&\multicolumn{1}{c}{$WWZ$}
&\multicolumn{1}{c}{HISZ scenario}\\
\tableline
 & & & \\ [-6.mm]
$\Delta g_1^{Z0}$ & -- & $\matrix{+1.00 \crc -0.53}$ & -- \\ [5.mm]
$\Delta\kappa^0_V$ & $\matrix{+0.64 \crc -0.35}$ & $\matrix{+0.32
\crc -0.22}$ & $\matrix{+0.43 \crc -0.19}$ \\ [5.mm]
$\lambda^0_V$ & $\matrix{+0.25 \crc -0.20}$ & $\matrix{+0.13
\crc -0.14}$ & $\matrix{+0.096 \crc -0.086}$ \\ [5.mm]
\end{tabular}
\end{table}
\newpage
\begin{table}
\caption{Sensitivities achievable at the 95\% confidence
level (CL) for anomalous $WWV$ couplings ($V=\gamma,\, Z$) in
$\protect{pp\rightarrow W^+W^- + 0~{\rm jet}\rightarrow \ell_1^+ \ell_2^-
p\llap/_T+0}$~jet, $\ell_{1,2}=e,\,\mu$, at NLO for the LHC,
including the residual background from $t\bar t$
production. Limits are shown for the case where only the
$WW\gamma$ or $WWZ$ couplings are allowed deviate from their SM values,
and for HISZ scenario where we assume $\Delta\kappa_\gamma$ and
$\lambda_\gamma$ as the independent couplings [see
Eqs.~(\protect{\ref{EQ:HISZ1}}) --~(\protect{\ref{EQ:HISZ3}})].
Interference effects between those couplings which are varied are fully
taken into account.
For the form factors we use the form of Eq.~(\protect{\ref{EQ:AFORM}})
with $n=2$. The transverse momentum threshold
for the jet veto and the $p_T(h)$ cut is taken to be 30~GeV for
$\int\!{\cal L}dt=10$~fb$^{-1}$, and 50~GeV for 100~fb$^{-1}$. The
$t\bar t$ cross section is calculated at LO with $m_t=176$~GeV.
The cuts summarized in Sec.~IIIB are imposed. \protect{\\ [3.mm]} }
\label{TABLE3}
\begin{tabular}{cccc}
 & & & \\[-6.mm]
\multicolumn{4}{c}{a) LHC, $\int\!{\cal L}dt=10$~fb$^{-1}$,
$\Lambda_{FF}=1$~TeV}\\
\multicolumn{1}{c}{coupling}
&\multicolumn{1}{c}{$WW\gamma$}
&\multicolumn{1}{c}{$WWZ$}
&\multicolumn{1}{c}{HISZ scenario}\\
\tableline
 & & & \\ [-6.mm]
$\Delta g_1^{Z0}$ & -- & $\matrix{+0.81 \crc -0.50}$ & -- \\ [5.mm]
$\Delta\kappa^0_V$ & $\matrix{+0.43 \crc -0.25}$ & $\matrix{+0.19
\crc -0.12}$ & $\matrix{+0.27 \crc -0.14}$ \\ [5.mm]
$\lambda^0_V$ & $\matrix{+0.15 \crc -0.14}$ & $\matrix{+0.063
\crc -0.071}$ & $\matrix{+0.052 \crc -0.049}$ \\ [5.mm]
\tableline
\tableline
& & & \\[-6.mm]
\multicolumn{4}{c}{b) LHC, $\int\!{\cal L}dt=100$~fb$^{-1}$,
$\Lambda_{FF}=1$~TeV (3~TeV)}\\
\multicolumn{1}{c}{coupling}
&\multicolumn{1}{c}{$WW\gamma$}
&\multicolumn{1}{c}{$WWZ$}
&\multicolumn{1}{c}{HISZ scenario}\\
\tableline
 & & & \\ [-6.mm]
$\Delta g_1^{Z0}$ & -- (--)& $\matrix{+0.62 \crc -0.50}$ $\left (\matrix{+0.22
\crc -0.17}\right )$ & -- (--)\\ [5.mm]
$\Delta\kappa^0_V$ & $\matrix{+0.31 \crc -0.18}$ $\left (\matrix{+0.067
\crc -0.040}\right)$ & $\matrix{+0.133 \crc -0.085}$ $\left(\matrix{+0.027
\crc -0.018}\right )$ & $\matrix{+0.201 \crc -0.110}$ $\left
(\matrix{+0.047 \crc -0.025}\right )$ \\ [5.mm]
$\lambda^0_V$ & $\matrix{+0.092 \crc -0.086}$ $\left (\matrix{+0.022
\crc -0.022}\right )$ & $\matrix{+0.042 \crc -0.040}$ $\left (\matrix{+0.0084
\crc -0.0111}\right )$ & $\matrix{+0.029 \crc -0.036}$ $\left
(\matrix{+0.0078 \crc -0.0079}\right )$ \\ [5.mm]
\end{tabular}
\end{table}
\newpage
%
%
\begin{figure}
\baselineskip 19pt
FIG. 1. Feynman rule for the general $WWV$ ($V= \gamma, Z$) vertex.
The factor $g_{WWV}$ is the vertex coupling strength:
$g_{WW\gamma}^{} = e$ and $g_{WWZ}^{} = e \cot\theta_{\rm W}$.
The vertex function $\Gamma_{\beta \mu \nu}(k,k_1,k_2)$ is given in
Eq.~(\protect{\ref{EQ:NSMCOUPLINGS}}).
\label{FIG:VERTEX}
\end{figure}
%
\begin{figure}
\baselineskip 19pt
FIG. 2. The inclusive differential cross section for the electron
transverse momentum in the reaction
$\protect{p \bar p \to W^+W^- + X \to e^+e^-p\llap/_T + X}$
at $\protect{\sqrt{s} = 1.8}$~TeV; a) in the Born approximation and b)
including NLO QCD corrections.
The curves are for the SM (solid lines), $\lambda^0_\gamma = -0.5$
(short dashed lines), $\Delta\kappa^0_\gamma = -0.5$ (short dotted
lines), $\lambda^0_Z=-0.5$ (long dashed lines), $\Delta\kappa^0_Z=-0.5$
(long dotted lines), and $\Delta g_1^{Z0} =
-1.0$ (dot-dashed lines). The cuts imposed are summarized in Sec.~IIIB.
\label{FIG:TWO}
\end{figure}
%
\begin{figure}
\baselineskip 19pt
FIG. 3. The inclusive differential cross section for the electron
transverse momentum in the reaction
$\protect{pp \to W^+W^- + X \to e^+e^-p\llap/_T + X}$
at $\protect{\sqrt{s} = 14}$~TeV; a) in the Born approximation and b)
including NLO QCD corrections.
The curves are for the SM (solid lines), $\lambda^0_\gamma = -0.25$
(short dashed lines), $\Delta\kappa^0_\gamma = -0.25$ (short dotted
lines), $\lambda^0_Z=-0.25$ (long dashed lines), $\Delta\kappa^0_Z=-0.25$
(long dotted lines), and $\Delta g_1^{Z0} =
-1.0$ (dot-dashed lines). The cuts imposed are summarized in Sec.~IIIB.
\label{FIG:THREE}
\end{figure}
%
\begin{figure}
\baselineskip 19pt
FIG. 4. The inclusive differential cross section for the missing
transverse momentum in the reaction
$\protect{p \bar p \to W^+W^- + X \to e^+e^-p\llap/_T + X}$
at $\protect{\sqrt{s} = 1.8}$~TeV; a) in the Born approximation and b)
including NLO QCD corrections.
The curves are for the SM (solid lines), $\lambda^0_\gamma = -0.5$
(short dashed lines), $\Delta\kappa^0_\gamma = -0.5$ (short dotted
lines), $\lambda^0_Z=-0.5$ (long dashed lines), $\Delta\kappa^0_Z=-0.5$
(long dotted lines), and $\Delta g_1^{Z0} =
-1.0$ (dot-dashed lines). The cuts imposed are summarized in Sec.~IIIB.
\label{FIG:FOUR}
\end{figure}
%
\begin{figure}
\baselineskip 19pt
FIG. 5. The inclusive differential cross section for the missing
transverse momentum in the reaction
$\protect{pp \to W^+W^- + X \to e^+e^-p\llap/_T + X}$
at $\protect{\sqrt{s} = 14}$~TeV; a) in the Born approximation and b)
including NLO QCD corrections.
The curves are for the SM (solid lines), $\lambda^0_\gamma = -0.25$
(short dashed lines), $\Delta\kappa^0_\gamma = -0.25$ (short dotted
lines), $\lambda^0_Z=-0.25$ (long dashed lines), $\Delta\kappa^0_Z=-0.25$
(long dotted lines), and $\Delta g_1^{Z0} =
-1.0$ (dot-dashed lines). The cuts imposed are summarized in Sec.~IIIB.
\label{FIG:FIVE}
\end{figure}
%
\begin{figure}
\baselineskip 19pt
FIG. 6. Ratio of the NLO and LO differential cross sections of the
missing transverse momentum (solid lines) and the transverse momentum of
the charged lepton (dashed lines) in the SM as a function of $p_T$ for a)
$\protect{p \bar p \to W^+W^- + X \to e^+e^-p\llap/_T + X}$ at
$\protect{\sqrt{s} = 1.8}$~TeV,
and b) $\protect{pp \to W^+W^- + X \to e^+e^-p\llap/_T + X}$
at $\protect{\sqrt{s} = 14}$~TeV. The cuts imposed are summarized in
Sec.~IIIB.
\label{FIG:SIX}
\end{figure}
%
\begin{figure}
\baselineskip 19pt
FIG. 7. The inclusive differential cross section for the
transverse momentum of the charged lepton pair in the reaction
$\protect{p \bar p \to W^+W^- + X \to e^+e^-p\llap/_T + X}$
at $\protect{\sqrt{s} = 1.8}$~TeV; a) in the Born approximation and b)
including NLO QCD corrections.
The curves are for the SM (solid lines), $\lambda^0_\gamma = -0.5$
(short dashed lines), $\Delta\kappa^0_\gamma = -0.5$ (short dotted
lines), $\lambda^0_Z=-0.5$ (long dashed lines), $\Delta\kappa^0_Z=-0.5$
(long dotted lines), and $\Delta g_1^{Z0} =
-1.0$ (dot-dashed lines). The cuts imposed are summarized in Sec.~IIIB.
\label{FIG:SEVEN}
\end{figure}
%
\begin{figure}
\baselineskip 19pt
FIG. 8. The inclusive differential cross section for the
transverse momentum of the charged lepton pair in the reaction
$\protect{pp \to W^+W^- + X \to e^+e^-p\llap/_T + X}$
at $\protect{\sqrt{s} = 14}$~TeV; a) in the Born approximation and b)
including NLO QCD corrections.
The curves are for the SM (solid lines), $\lambda^0_\gamma = -0.25$
(short dashed lines), $\Delta\kappa^0_\gamma = -0.25$ (short dotted
lines), $\lambda^0_Z=-0.25$ (long dashed lines), $\Delta\kappa^0_Z=-0.25$
(long dotted lines), and $\Delta g_1^{Z0} =
-1.0$ (dot-dashed lines). The cuts imposed are summarized in Sec.~IIIB.
\label{FIG:EIGHT}
\end{figure}
%
\begin{figure}
\baselineskip 19pt
FIG. 9. The $p_T(e^+e^-)$ differential cross section for a)
$\protect{p \bar p \to W^+W^- + X \to e^+e^-p\llap/_T + X}$
at $\protect{\sqrt{s} = 1.8}$~TeV, and b) $\protect{pp \to W^+W^- + X
\to e^+e^-p\llap/_T + X}$ at $\protect{\sqrt{s} = 14}$~TeV. The
inclusive NLO differential cross section (solid line) is
decomposed into the \protect{${\cal O}(\alpha_s)$} \protect{0-jet}
(dotted line) and LO \protect{1-jet} (dot dashed line) exclusive
differential cross sections.  For comparison, the Born cross section
(dashed line) is also shown.
The cuts imposed are summarized in Sec.~IIIB. For the jet definitions,
we have used Eqs.~(\protect{\ref{EQ:TEVJET}})
and~(\protect{\ref{EQ:LHCJET}}).
\label{FIG:NINE}
\end{figure}
%
\begin{figure}
\baselineskip 19pt
FIG. 10. The LO differential cross section for the $e^+e^-$ transverse
momentum for a) $\protect{p \bar p \to e^+e^-p\llap/_T + X}$
at $\protect{\sqrt{s} = 1.8}$~TeV, and b) $\protect{pp
\to e^+e^-p\llap/_T + X}$ at $\protect{\sqrt{s} = 14}$~TeV. The SM
$W^+W^-$ cross section (solid line) is shown, together with the $t\bar
t\to W^+W^-b\bar b\to e^+e^-p\llap/_T + X$ rate for $m_t=176$~GeV
(dashed line), the $ZZ\to
e^+e^-p\llap/_T + X$ cross section (dot dashed line), the $W^\pm Z\to
e^+e^-p\llap/_T + X$ cross section where one of the two like sign
charged leptons is produced with a rapidity outside the range covered by
the detector (long dashed line), and the $Z+1~{\rm jet}\to
e^+e^-p\llap/_T + X$ rate, where the jet disappears through the beam
hole (dotted line). The cuts imposed are summarized in Secs.~IIIB
and~IIID.
\label{FIG:TEN}
\end{figure}
%
\begin{figure}
\baselineskip 19pt
FIG. 11. The LO differential cross section for the $e^+e^-$ transverse
momentum for a) $\protect{p \bar p \to e^+e^-p\llap/_T + X}$
at $\protect{\sqrt{s} = 1.8}$~TeV, and b) $\protect{pp
\to e^+e^-p\llap/_T + X}$ at $\protect{\sqrt{s} = 14}$~TeV. We show the SM
$W^+W^-$ cross section with (dashed lines) and without an $|m(e^+e^-)
-M_Z|>10$~GeV cut (solid line). The additional cuts imposed are summarized in
Sec.~IIIB.
\label{FIG:ELEVEN}
\end{figure}
%
\begin{figure}
\baselineskip 19pt
FIG. 12. The LO differential cross section for the $e^+e^-$ transverse
momentum for a) $\protect{p \bar p \to t\bar t\to e^+e^-p\llap/_T + X}$
at $\protect{\sqrt{s} = 1.8}$~TeV, and b) $\protect{pp \to t\bar t
\to e^+e^-p\llap/_T + X}$ at $\protect{\sqrt{s} = 14}$~TeV. The solid
lines show the inclusive differential cross section. The dashed, dotted,
and dot-dashed lines give the 0-jet, 1-jet, and 2-jet exclusive cross
sections, respectively. The long-dashed curves show the $p_T(e^+e^-)$
distribution with a cut on the total transverse momentum of the hadrons
in the event of $p_T(h)<20$~GeV (50~GeV) at the Tevatron (LHC) [see
Eq.~(\protect{\ref{EQ:PTHAD}})]. We assume a top quark mass of $m_t=176$~GeV.
The cuts imposed are summarized in Sec.~IIIB. For the jet definitions,
we have used Eqs.~(\protect{\ref{EQ:TEVJET}})
and~(\protect{\ref{EQ:LHCJET}}).
\label{FIG:TWELVE}
\end{figure}
%
\begin{figure}
\baselineskip 19pt
FIG. 13. \sloppy{The $e^+e^-$ transverse momentum distribution for
$\protect{p\bar p\to}$ $W^+W^-+0~{\rm jet}\to
e^+e^-p\llap/_T+0~{\rm jet}$ at ${\cal O}(\alpha_s)$ (solid line),
$\protect{p\bar p\to t\bar t\to e^+e^-p\llap/_T+0}$~jet (dashed line),
and $\protect{p\bar p\to t\bar t\to e^+e^-p\llap/_T+X}$ with the
indicated $p_T(h)$ cut imposed (dotted line), at the Tevatron. In part a) a
jet-defining transverse momentum threshold of $p_T(j)>20$~GeV is used;
in part b) the threshold is lowered to $p_T(j)
>10$~GeV. For $W^+W^-$ production at ${\cal O}(\alpha_s)$, a jet veto
and a $p_T(h)$ cut are equivalent. The additional cuts imposed are
summarized in Sec.~IIIB. }
\label{FIG:THIRTEEN}
\end{figure}
%
\begin{figure}
\baselineskip 19pt
FIG. 14. \sloppy{The $e^+e^-$ transverse momentum distribution for
$\protect{pp\to}$ $W^+W^-+0~{\rm jet}\to
e^+e^-p\llap/_T+0~{\rm jet}$ at ${\cal O}(\alpha_s)$ (solid line),
$\protect{pp\to t\bar t}$ $\to e^+e^-p\llap/_T+0$~jet (dashed line),
and $\protect{pp\to t\bar t}$ $\to e^+e^-p\llap/_T+X$ with the
indicated $p_T(h)$ cut imposed (dotted line), at the LHC. In part a) a
jet-defining transverse momentum threshold of $p_T(j)>50$~GeV is used;
in part b) the threshold is lowered to $p_T(j)
>30$~GeV. For $W^+W^-$ production at ${\cal O}(\alpha_s)$, a jet veto
and a $p_T(h)$ cut are equivalent. The additional cuts imposed are
summarized in Sec.~IIIB. }
\label{FIG:FOURTEEN}
\end{figure}
%
\begin{figure}
\baselineskip 19pt
FIG. 15. The combined differential cross section for the $e^+e^-$
transverse momentum from $W^+W^-\to e^+e^-p\llap/_T + 0$~jet and
$t\bar t\to e^+e^-p\llap/_T + 0$~jet for a) $\protect{p \bar p}$
collisions at $\protect{\sqrt{s} = 1.8}$~TeV, and b) $\protect{pp}$
collisions at $\protect{\sqrt{s} = 14}$~TeV. The curves are for the SM
(solid line), and two sets of anomalous couplings in the HISZ scenario
[Eqs.~(\protect{\ref{EQ:HISZ1}}) --~(\protect{\ref{EQ:HISZ3}})]. The
dashed line shows the result for ($\lambda_\gamma^0=-0.5$,
$\Delta\kappa_\gamma^0=0$) [($\lambda_\gamma^0=-0.25$,
$\Delta\kappa_\gamma^0=0$)] at the Tevatron [LHC]. The dotted line
corresponds to ($\lambda_\gamma^0=0$, $\Delta\kappa_\gamma^0=-0.5$)
[($\lambda_\gamma^0=0$, $\Delta\kappa_\gamma^0=-0.25$)].
The cuts imposed are summarized in Sec.~IIIB. For the jet definitions,
we have used Eqs.~(\protect{\ref{EQ:TEVJET}})
and~(\protect{\ref{EQ:LHCJET}}). A top quark mass of $m_t=176$~GeV
was used.
\label{FIG:FIFTEEN}
\end{figure}
%
\begin{figure}
\baselineskip 19pt
FIG. 16. The LO $e^+e^-$ transverse momentum distribution for a)
$\protect{p \bar p \to W^+W^-\to e^+e^-p\llap/_T}$ at
$\protect{\sqrt{s} = 1.8}$~TeV, and b) $\protect{pp \to W^+W^-\to
e^+e^-p\llap/_T}$ at $\protect{\sqrt{s} = 14}$~TeV. The solid lines show
the result for the direct $W\to e\nu$ decays. The dashed (dotted)
lines represents the differential cross sections if one (both) charged
leptons in the final state originate from $W\to\tau\nu_\tau\to
e\nu_e\nu_\tau\bar\nu_\tau$. The cuts imposed are summarized in
Sec.~IIIB.
\label{FIG:FIFTEENA}
\end{figure}
%
\begin{figure}
\baselineskip 19pt
FIG. 17. Limit contours at the \protect{95\%~CL} for
\protect{$p\bar p\rightarrow
W^+W^-+X\rightarrow\ell_1^+\ell_2^-p\llap/_T+X$}, $\ell_{1,2}=e,\,\mu$,
derived from the $p_T(\ell_1^+\ell_2^-)$ distribution at the Tevatron for
$\int\!{\cal L}dt=1$~fb$^{-1}$. Contours are shown in three planes:
a) the \protect{$\Delta\kappa^0_Z$} \protect{--}~\protect{$\lambda^0_Z$}
plane,
b) the \protect{$\Delta\kappa^0_Z$} \protect{--}~\protect{$\Delta
g_1^{Z0}$} plane, and c) the \protect{$\Delta g_1^{Z0}$}
\protect{--}~\protect{$\lambda^0_Z$}
plane. The solid lines give the results for LO $W^+W^-$ production,
ignoring the $t\bar t$ background. The dashed lines
show the limits which are obtained if the top quark background is taken
into account and the inclusive NLO $W^+W^-$ cross section is used. The
dotted lines, finally, display the bounds which are achieved from the
exclusive NLO $W^+W^-+0$~jet channel, including the residual $t\bar
t\to W^+W^-+0$~jet background.
The cuts imposed are summarized in Sec.~IIIB. For the top quark mass we
assume $m_t=176$~GeV, and for the jet definition,
we have used Eq.~(\protect{\ref{EQ:TEVJET}}).
\label{FIG:SIXTEEN}
\end{figure}
%
\begin{figure}
\baselineskip 19pt
FIG. 18. Limit contours at the \protect{95\%~CL} for \protect{$pp\rightarrow
W^+W^-+X\rightarrow\ell_1^+\ell_2^-p\llap/_T+X$}, $\ell_{1,2}=e,\,\mu$,
derived from the $p_T(\ell_1^+\ell_2^-)$ distribution at the LHC for
$\int\!{\cal L}dt=10$~fb$^{-1}$. Contours are shown in three planes:
a) the \protect{$\Delta\kappa^0_Z$}
\protect{--}~\protect{$\lambda^0_Z$} plane,
b) the \protect{$\Delta\kappa^0_Z$} \protect{--}~\protect{$\Delta
g_1^{Z0}$} plane, and c) the \protect{$\Delta g_1^{Z0}$}
\protect{--}~\protect{$\lambda^0_Z$}
plane. The solid lines give the results for LO $W^+W^-$ production,
ignoring the $t\bar t$ background. The dashed lines
show the limits which are obtained if the top quark background is taken
into account and the inclusive NLO $W^+W^-$ cross section is used. The
dotted lines, finally, display the bounds which are achieved from the
exclusive NLO $W^+W^-+0$~jet channel, including the residual $t\bar
t\to W^+W^-+0$~jet background.
The cuts imposed are summarized in Sec.~IIIB. For the top quark mass we
assume $m_t=176$~GeV, and for the jet definition,
we have used Eq.~(\protect{\ref{EQ:LHCJET}}).
\label{FIG:SEVENTEEN}
\end{figure}
%
\begin{figure}
\baselineskip 19pt
FIG. 19. Limit contours at the \protect{95\%~CL}, derived from the NLO
$p_T(\ell_1^+\ell_2^-)$, $\ell_{1,2}=e,\,\mu$, distribution, for a)
\protect{$p\bar p\rightarrow
W^+W^-+0~{\rm jet}\rightarrow\ell_1^+\ell_2^-p\llap/_T+0$~jet}
at $\protect{\sqrt{s} = 1.8}$~TeV with
$\int\!{\cal L}dt=1$~fb$^{-1}$, and b) \protect{$pp\rightarrow
W^+W^-+0~{\rm jet}\rightarrow\ell_1^+\ell_2^-p\llap/_T+0$~jet}
at $\protect{\sqrt{s} = 14}$~TeV with
$\int\!{\cal L}dt=10$~fb$^{-1}$ in the \protect{$\Delta\kappa^0_V$}
\protect{--}~\protect{$\lambda^0_V$} plane. The solid line displays the
limits which are achieved if \protect{$\Delta\kappa^0_Z$} and
\protect{$\lambda^0_Z$} only are allowed to deviate from their SM
values. The dotted and dashed lines show the results obtained in the
HISZ scenario [see Eqs.~(\protect{\ref{EQ:HISZ1}})
--~(\protect{\ref{EQ:HISZ3}})] and by varying the $WW\gamma$ couplings only.
The effect of the residual $t\bar t\to e^+e^-p\llap/_T+0$~jet background
is included in the contours shown.
The cuts imposed are summarized in Sec.~IIIB. For the top quark mass we
assume $m_t=176$~GeV, and for the jet definition,
we have used Eqs.~(\protect{\ref{EQ:TEVJET}})
and~(\protect{\ref{EQ:LHCJET}}).
\label{FIG:EIGHTEEN}
\end{figure}
%
\begin{figure}
\baselineskip 19pt
FIG. 20. Limit contours at the \protect{95\%~CL}, derived from the NLO
$p_T(\ell_1^+\ell_2^-)$, $\ell_{1,2}=e,\,\mu$, distribution, for a)
\protect{$p\bar p\rightarrow
W^+W^-+0~{\rm jet}\rightarrow\ell_1^+\ell_2^-p\llap/_T+0$~jet} at
$\protect{\sqrt{s} = 1.8}$~TeV, and b) \protect{$pp\rightarrow
W^+W^-+0~{\rm jet}\rightarrow\ell_1^+\ell_2^-p\llap/_T+0$~jet}
at $\protect{\sqrt{s} = 14}$~TeV in the HISZ scenario
[see Eqs.~(\protect{\ref{EQ:HISZ1}}) --~(\protect{\ref{EQ:HISZ3}})]. In
part a) the solid and dashed lines give the limits for integrated
luminosities of $\int\!{\cal L}dt=1$~fb$^{-1}$ and
10~fb$^{-1}$, respectively. The form factor scale in
both cases is $\Lambda_{FF}=1$~TeV. In part b) results are displayed for
$\int\!{\cal L}dt=10$~fb$^{-1}$ (solid curve) and
$\int\!{\cal L}dt=100$~fb$^{-1}$ (dashed curve) with
$\Lambda_{FF}=1$~TeV, and $\int\!{\cal L}dt=100$~fb$^{-1}$ with
$\Lambda_{FF}=3$~TeV (dotted curve). The effect of the residual $t\bar
t\to e^+e^-p\llap/_T+0$~jet background is included in the contours shown.
The cuts imposed are summarized in Sec.~IIIB. For the
top quark mass we assume $m_t=176$~GeV. The jet definition criteria are
described in Sec.~IIIE.
\label{FIG:NINETEEN}
\end{figure}
%
\begin{figure}
\baselineskip 19pt
FIG. 21. Comparison of the expected sensitivities on anomalous $WWV$
couplings in the HISZ scenario [see Eqs.~(\protect{\ref{EQ:HISZ1}})
--~(\protect{\ref{EQ:HISZ3}})] from \protect{$e^+e^-\to
W^+W^-\to\ell\nu jj$} at \protect{LEP~II} ($\protect{\sqrt{s}=190}$~GeV,
$\int\!{\cal L}dt=500$~pb$^{-1}$), and di-boson production processes at the
Tevatron (\protect{$\int\!{\cal L}dt=10$~fb$^{-1}$}). Except for the
short dashed curve, which shows the result for \protect{$p\bar p\rightarrow
W^+W^-+0~{\rm jet}\rightarrow\ell_1^+\ell_2^-p\llap/_T+0$~jet} at
$\protect{\sqrt{s} = 1.8}$~TeV, all curves are taken from
Ref.~\protect{\cite{DPF}}.
\label{FIG:TWENTY}
\end{figure}
%
%

\begin{references}
%
\bibitem{LAGRANGIAN}
K.~Hagiwara, R.D.~Peccei, D.~Zeppenfeld, and K.~Hikasa,
Nucl. Phys. {\bf B~282}, 253 (1987).
%
\bibitem{Ber}
F.~Berends and A.~van Sighem, INLO-PUB-7-95
(preprint, June~1995) and references therein.
%
\bibitem{Brenna}
B.~Flaugher, FERMILAB-Conf-95/180-E (preprint, June~1995), to appear in the
Proceedings of the {\it ``Les Recontres de Physique de la Vallee d'
Aoste''}, La Thuile, Italy, March~5 --~11, 1995.
%
\bibitem{FIRSTWW}
R.W.~Brown and K.O.~Mikaelian,
Phys. Rev. {\bf D~19}, 922 (1979).	
%
\bibitem{BS}
C.L.~Bilchak and J.~Stroughair, Z.~Phys. {\bf C~23}, 377 (1984).
%
\bibitem{WWAC}
M.~Kuroda {\it et al.}, Nucl. Phys. {\bf B~284}, 271 (1987);
C.-H.~Chang and S.-C.~Lee, Phys. Rev. {\bf D~37}, 101 (1988).
%
\bibitem{HWZ}
K.~Hagiwara, J.~Woodside, and D.~Zeppenfeld,
Phys. Rev. {\bf D~41}, 2113 (1990).	
%
\bibitem{EHLQ}
E.~Eichten, I.~Hinchliffe, K.~Lane, and C.~Quigg,
Rev. Mod. Phys. {\bf 56}, 579 (1984); {\bf 58}, 1065(E) (1986).  
%
\bibitem{CDFWW}
F.~Abe {\it et al.} (CDF Collaboration), FERMILAB-Pub-95/036-E
(preprint, March 1995), to appear in Phys. Rev. Lett.
%
\bibitem{DOWW}
S.~Abachi {\it et al.} (D\O\ Collaboration), FERMILAB-Pub-95/044-E
(preprint, March 1995), to appear in Phys. Rev. Lett.
%
\bibitem{WW}
J.~Ohnemus, Phys. Rev. {\bf D~44}, 1403 (1991).	
%
\bibitem{WWFRIX}
S.~Frixione,
Nucl. Phys. {\bf B~410}, 280 (1993).  
%
\bibitem{Doug}
T.~Diehl, FERMILAB-Conf-95/165-E (preprint, July~1995), to appear in the
Proceedings of the {\sl ``10th Topical Workshop on Proton
Antiproton Physics''}, Fermilab, May~1995.
%
\bibitem{GPJ}
D.~Amidei {\it et al.}, FERMILAB-Conf-94/249-E (preprint, August 1994),
to appear in the Proceedings of the {\it DPF'94
Conference}, Albuquerque, New Mexico, August~2 --~6, 1994;
S.~Holmes, G.~Dugan, and S.~Peggs, {\it Proceedings of the 1990
Summer Study on High Energy Physics: Research Directions for the
Decade}, Snowmass, Colorado, edited by E.~L.~Berger, p.~674.
%
\bibitem{LHC}
The LHC Study Group, Design Study of the Large Hadron Collider,
CERN 91-03, 1991; L.~R.~Evans, Proceedings of the {\it ``27th
International Conference on High
Energy Physics''}, Glasgow, Scotland, July 1994, Vol.~II, p.~1417 and
CERN-AC-95-002 (preprint, June~1995);
C.~W.~Fabjan, CERN-PPE/95-25 (preprint, February 1995).
%
\bibitem{NLOMC}
H.~Baer, J.~Ohnemus, and J.F.~Owens,
Phys. Rev. {\bf D~40}, 2844 (1989);  	
H.~Baer, J.~Ohnemus, and J.F.~Owens,
Phys. Rev. {\bf D~42}, 61 (1990);
Phys. Lett. {\bf B~234}, 127 (1990);	
J.~Ohnemus and J.F.~Owens,
Phys. Rev. {\bf D~43}, 3626 (1991);	
J.~Ohnemus,
Phys. Rev. {\bf D~44}, 1403 (1991);	
J.~Ohnemus,
Phys. Rev. {\bf D~44}, 3477 (1991);	
H.~Baer and M.H.~Reno,
Phys. Rev. {\bf D~43}, 2892 (1991);	
B.~Bailey, J.~Ohnemus, and J.F.~Owens,
Phys. Rev. {\bf D~46}, 2018 (1992);	
J.~Ohnemus,
Phys. Rev. {\bf D~47}, 940 (1993); 	
J.~Ohnemus and W.J.~Stirling,
Phys. Rev. {\bf D~47}, 2722 (1993);  	
H.~Baer, B.~Bailey, and J.F.~Owens,
Phys. Rev. {\bf D~47}, 2730 (1993);	
L.~Bergmann,
Ph.D. dissertation, Florida State University,
report No. FSU-HEP-890215, 1989 (unpublished).
%
\bibitem{HEL}
C.~Bilchak, R.~Brown, and J.~Stroughair, Phys. Rev. {\bf D~29}, 375
(1984).
%
\bibitem{NEWJO}
J.~Ohnemus, Phys. Rev. {\bf D~50}, 1931 (1994). 
%
\bibitem{GAUGE}
U.~Baur and D.~Zeppenfeld, MAD/PH/878 (preprint, March~1995), to appear
in Phys. Rev. Lett.;
E.N.~Argyres {\it et al.}, INLO-PUB-8/95 (preprint, July~1995);
C.~Papadopoulos, Phys. Lett. {\bf B~352}, 144 (1995);
J.~Papavassiliou and A.~Pilaftsis, NYU-TH/95-02 (preprint, June~1995)
and NYU-TH/95-03 (preprint, July~1995);
G.~Lopez Castro, J.L.M.~Lucio, and J.~Pestieau,
Mod.~Phys.~Lett. {\bf A~6}, 3679 (1991) and hep-ph/9504351 (preprint,
April~1995);
M.~Nowakowski and A.~Pilaftsis, Z.~Phys. {\bf C~60}, 121 (1993);
A.~Aeppli, F.~Cuypers, and G.~J. van Oldenborgh, Phys. Lett. {\bf B~314},
413 (1993).
%
\bibitem{DIMREG}
G.'t Hooft and M.~Veltman,
Nucl. Phys. {\bf B~44}, 189 (1972).  
%
\bibitem{VVJET}
U.~Baur, E.W.N.~Glover, and J.J. van der Bij,
Nucl. Phys. {\bf B~318}, 106 (1989);
V.~Barger, T.~Han, J.~Ohnemus, and D.~Zeppenfeld,
Phys. Rev. {\bf D~41}, 2782 (1990).  
%
\bibitem{DPF}
H.~Aihara {\it et al.}, FERMILAB-Pub-95/031 (preprint, March~1995), to
appear in {\sl ``Electroweak Symmetry Breaking and Beyond the Standard
Model''}, eds. T.~Barklow, S.~Dawson, H.~Haber and J.~Siegrist.
%
\bibitem{CDFWG}
F.~Abe {\it et al.} (CDF Collaboration), Phys. Rev. Lett. {\bf 74},
1936 (1995).
%
\bibitem{DOWG}
S.~Abachi {\it et al.} (D\O\ Collaboration), FERMILAB-Pub-95/101-E (preprint,
May~1995), to appear in Phys. Rev. Lett.
%
\bibitem{DE}
A.~De Rujula {\it et al.}, Nucl. Phys. {\bf B~384}, 31 (1992);
P.~Hern\'andez and F.J.~Vegas, Phys. Lett. {\bf B~307}, 116 (1993).
%
\bibitem{BL}
C.~Burgess and D.~London, Phys. Rev. Lett. {\bf 69}, 3428 (1992) and
Phys. Rev. {\bf D~48}, 4337 (1993).
%
\bibitem{HISZ}
K.~Hagiwara, S.~Ishihara, R.~Szalapski, and D.~Zeppenfeld,
Phys. Lett. {\bf B~283}, 353 (1992) and Phys. Rev. {\bf D~48}, 2182 (1993).
%
\bibitem{DV}
S.~Dawson and G.~Valencia, Nucl. Phys. {\bf B~439}, 3 (1995).
%
\bibitem{MUON}
P.~M\'ery, S.E.~Moubarik, M.~Perrottet, and F.M.~Renard,
Z. Phys. {\bf C~46}, 229 (1990);
C.~Arzt, M.~Einhorn, and J.~Wudka, Phys. Rev. {\bf D~49}, 1370 (1994).
%
\bibitem{BDEC}
S.~P.~Chia, Phys. Lett. {\bf B~240}, 465 (1990);
K.~Numata, Z. Phys. {\bf C~52}, 691 (1991);
K.~A.~Peterson, Phys. Lett. {\bf B~282}, 207 (1992);
T.~G.~Rizzo, Phys. Lett. {\bf B~315}, 471 (1993);
U.~Baur, Proceedings of the {\it ``Workshop on B Physics at Hadron
Accelerators''}, Snowmass, Colorado, June 1993, p.~455;
X.~He and B.~McKellar, Phys. Lett. {\bf B~320}, 165 (1994).
R.~Martinez, M.~A.~P\'erez, and J.~J.~Toscano, Phys. Lett.
{\bf B~340}, 91 (1994).
%
\bibitem{CLEO}
M. S. Alam {\it et al.} (CLEO Collaboration), Phys. Rev. Lett. {\bf
74}, 2885 (1995).
%
\bibitem{HE}
X.~He, Phys. Lett. {\bf B~319}, 327 (1993).
%
\bibitem{BAL}
G.~Baillie, Z. Phys. {\bf C~61}, 667 (1994).
%
\bibitem{HMK}
X.~He and B.~McKellar, Phys. Rev. {\bf D~51}, 6484 (1995).
%
\bibitem{NOV}
G.~Belanger, F.~Boudjema, and D.~London, Phys. Rev. Lett. {\bf 65}, 2943
(1990);
O.~Eboli {\it et al.}, Phys. Lett. {\bf B~339}, 119
(1994); F.~M.~Renard and C.~Verzegnassi, Phys. Lett. {\bf
B~345}, 500 (1995).
%
\bibitem{HMHK}
K.~Hagiwara, S.~Matsumoto, D.~Haidt, and C.~S.~Kim,
Z.~Phys. {\bf C~64}, 559 (1994);
S.~Matsumoto, KEK-TH-418 (preprint, November 1994).
%
\bibitem{ARZT}
C. Arzt, M.B.~Einhorn, and J.~Wudka,  Nucl. Phys. {\bf B~433}, 41
(1994).
%
\bibitem{FLS}
A.~Falk, M.~Luke, and E.~Simmons, Nucl. Phys. {\bf B~365}, 523 (1991).
%
\bibitem{CHIRALP}
J.~Bagger, S.~Dawson, and G.~Valencia, Nucl. Phys. {\bf B~399}, 364 (1993);
M.J.~Herrero and E.~Ruiz Morales, Nucl. Phys. {\bf B~418}, 431 (1994);
J.~Wudka, Int.~J. Mod. Phys. {\bf A~9}, 2301 (1994).
%
\bibitem{FORM}
J.A.M.~Vermaseren, FORM User's Manual, NIKHEF-H, Amsterdam, 1989.
%
\bibitem{CORNWALL}
J.M.~Cornwall, D.N.~Levin, and G.~Tiktopoulos,
Phys. Rev. Lett. {\bf 30}, 1268 (1973);
Phys. Rev. {\bf D~10}, 1145 (1974);
C.H.~Llewellyn Smith, Phys. Lett. {\bf B~46}, 233 (1973);
S.D.~Joglekar, Ann. of Phys. {\bf 83}, 427 (1974).
%
\bibitem{FORMF}
U.~Baur and D.~Zeppenfeld,
Phys. Lett. {\bf B~201}, 383 (1988);
U.~Baur and D.~Zeppenfeld,
Nucl. Phys. {\bf B~308}, 127 (1988). 
%
\bibitem{BUCH}
W.~Buchm\"uller and D.~Wyler, Nucl. Phys. {\bf B~268}, 621 (1986).
%
\bibitem{BIL}
M.~Bilenky {\it et al.}, Nucl. Phys. {\bf B~409}, 22 (1993) and
references therein.
%
\bibitem{KAO}
D.~Dicus and C.~Kao, Phys. Rev. {\bf D~43}, 1555 (1991); C.~Kao, private
communication.
%
\bibitem{TOPMASS1}
F.~Abe {\it et~al.} (CDF Collaboration), Phys. Rev. Lett. {\bf 73}, 225
(1994); Phys. Rev. {\bf D~50}, 2966 (1994); Phys. Rev. Lett. {\bf 74},
2626 (1995) and FERMILAB-Pub-95/149-E (preprint, June~1995), submitted
to Phys. Rev. Lett.
%
\bibitem{TOPMASS2}
S.~Abachi {\it et~al.} (D\O\ Collaboration), Phys. Rev. Lett. {\bf 74},
2422 (1995); Phys. Rev. Lett. {\bf 74}, 2632 (1995) and
FERMILAB-Pub-95/020-E (preprint, May 1995), submitted to Phys. Rev. {\bf
D}.
%
\bibitem{LEP}
D.~Schaile, Proceedings of the
{\it ``27th International Conference on High Energy Physics''}, Glasgow,
Scotland, July~1994, Vol.~I, p.~27; The LEP Collaborations,
CERN-PPE/94-187 (preprint, November 1994).
%
\bibitem{SLC}
K.~Abe {\it et al.} (SLD Collaboration), Phys. Rev. Lett. {\bf 73}, 25
(1994).
%
\bibitem{MW}
J.~Alitti {\it et al.} (UA2 Collaboration),
Phys. Lett. {\bf B~241}, 150 (1990);
F.~Abe {\it et~al.} (CDF Collaboration),
Phys. Rev. Lett. {\bf 65}, 2243 (1990);
Phys. Rev. {\bf D~43}, 2070 (1991);
F.~Abe {\it et al.} (CDF Collaboration), Phys. Rev. Lett. {\bf 75}, 11
(1995) and FERMILAB-Pub-95/033-E (preprint, March 1995), to appear in
Phys. Rev. {\bf D}.
%
\bibitem{MSBAR}
W.A.~Bardeen, A.J.~Buras, D.W.~Duke, and T.~Muta,
Phys. Rev. {\bf D~18}, 3998 (1978).	
%
\bibitem{MRSA}
A.D.~Martin, R.G.~Roberts, and W.J.~Stirling,
Phys. Rev. {\bf D~50}, 6734 (1994).	
%
\bibitem{HERA}
M.~Derrick {\it et al.} (ZEUS Collaboration), Z.~Phys. {\bf C~65}, 379
(1995); T.~Ahmed {\it et al.} (H1~Collaboration), Nucl. Phys. {\bf
B~439}, 471 (1995).
%
\bibitem{ASYMM}
F.~Abe {\it et al.} (CDF Collaboration), Phys. Rev.
Lett. {\bf 74}, 850 (1995); P.~de Barbaro, FERMILAB-Conf-95/164-E
(preprint, June~1995), to appear in the Proceedings of the {\sl ``10th
Topical Workshop on Proton Antiproton Physics''}, Fermilab, May~1995, .
%
\bibitem{NA51}
A.~Baldit {\it et al.} (NA51 Collaboration), Phys. Lett. {\bf B~332},
244 (1994).
%
\bibitem{PILE}
C.~Albajar {\it et al.}, {\it Proceedings of the ECFA
Workshop on LHC Physics}, Aachen, FRG, 1990, Vol.~II, p.~621.
%
\bibitem{ATLAS}
D.~Gingrich {\it et al.} (ATLAS Collaboration), ATLAS Letter of
Intent, CERN-LHCC-92-4 (October~1992); W.~W.~Armstrong {\it et al.}
(ATLAS Collaboration), ATLAS Technical Design Report, CERN-LHCC-94-43
(December 1994).
%
\bibitem{CMS}
M.~Della Negra {\it et al.} (CMS Collaboration), CMS Letter of
Intent, CERN-LHCC-92-3 (October~1992); G.~L.~Bayatian {\it et al.}
(CMS Collaboration), CMS Technical Design Report, CERN-LHCC-94-38
(December 1994).
%
\bibitem{RCDF}
F.~Abe {\it et~al.} (CDF Collaboration), Phys. Rev. {\bf D~45}, 3921
(1992).
%
\bibitem{WVENHANC}
S.~Frixione, P.~Nason, and G.~Ridolfi,
Nucl. Phys. {\bf B~383}, 3 (1992);
%
\bibitem{BHO}
U.~Baur, T.~Han, and J.~Ohnemus, Phys. Rev. {\bf D~48}, 5140 (1993).
%
\bibitem{BHO1}
U.~Baur, T.~Han, and J.~Ohnemus, Phys. Rev. {\bf D~51}, 3381 (1995).
%
\bibitem{LHCJET}
G.~Ciapetti and A.~Di Ciaccio, {\it Proceedings of the ECFA
Workshop on LHC Physics}, Aachen, FRG, 1990, Vol.~II, p.~155.
%
\bibitem{RESUM}
T.~Han, R.~Meng, and J.~Ohnemus, Nucl. Phys. {\bf B~384}, 59 (1992);
P.~Arnold and R.~Kauffman, Nucl. Phys. {\bf B~349}, 381 (1991);
R.~Kauffman, Phys. Rev. {\bf D~44}, 1415 (1991); Phys. Rev. {\bf D~45},
1512 (1992);
C.P.~Yuan, Phys. Lett. {\bf B~283}, 395 (1992).
%
\bibitem{KS}
R.~Kleiss and W.~J.~Stirling, Z.~Phys. {\bf C~40}, 419 (1988);
V.~Barger, J.~Ohnemus, and R.J.N.~Phillips, Int. J.~Mod. Phys. {\bf A~4},
617 (1989).
%
\bibitem{CAL}
F.~Abe {\it et al.} (CDF Collaboration), Phys. Rev. {\bf D~45}, 2249
(1992); S.~Abachi {\it et al.} (D\O\ Collaboration), Nucl. Instrum.
Meth. {\bf A~338}, 185 (1994).
%
\bibitem{CAVA}
F.~Cavanna, D.~Denegri and T.~Rodrigo, {\it Proceedings of the ECFA
Workshop on LHC Physics}, Aachen, FRG, 1990, Vol.~II, p.~329.
%
\bibitem{LAD}
G.~Ladinsky and C.P. Yuan, Phys. Rev. {\bf D~43}, 789 (1991); C.P.~Yuan,
MSUHEP-50228 (preprint, February 1995); S.~Keller, private
communication.
%
\bibitem{PMS}
V.~Barger and R.J.N.~Phillips, Phys. Rev. Lett. {\bf 55}, 2752 (1985);
H.~Baer, {\it et al.}, Phys. Rev. {\bf D~37}, 3152 (1988).
%
\bibitem{YUAN}
S.~Willenbrock and D.~Dicus, Phys.  Rev.  {\bf D~34}, 155 (1986);
C.-P.~Yuan,   Phys.  Rev. {\bf D~41},  42 (1990);
R.~K.~Ellis and S.~Parke, Phys.  Rev. {\bf D~46},  3785  (1992);
D.~Carlson and C.-P.~Yuan, Phys. Lett. {\bf B~306}, 386 (1993);
G.~Bordes and B.~van~Eijk, Nucl. Phys. {\bf B~435},  23  (1995).
%
\bibitem{ALAN}
W.~Marciano, A.~Stange, and S.~Willenbrock, Phys. Rev. {\bf D~50}, 4491
(1994).
%
\bibitem{CORT}
S.~Cortese and R.~Petronzio, Phys. Lett. {\bf B~253}, 494 (1991);
T.~Stelzer and S.~Willenbrock, DTP/95/40 (preprint, May~1995).
%
\bibitem{CHARM}
U.~Baur, F.~Halzen, S.~Keller, M.L.~Mangano, and K.~Riesselmann, Phys.
Lett. {\bf B~318}, 544 (1993).
%
\bibitem{MANG}
M.~Mangano, Nucl. Phys. {\bf B~405}, 536 (1993);
P.J.~Rijken and W.L.~van Neerven, Phys. Rev. {\bf D~52}, 149 (1995).
%
\bibitem{HTOP}
H.~Baer, V.~Barger, and R.J.N.~Phillips, Phys. Rev. {\bf D~39}, 3310
(1989); Phys. Rev. {\bf D~39}, 2809 (1989); R.~Kleiss, A.D.~Martin, and
W.J.~Stirling, Z.~Phys. {\bf C~39}, 393 (1988); S.~Gupta and D.P.~Roy,
Z.~Phys. {\bf C~39}, 417 (1988).
%
\bibitem{NLOTOP}
P.~Nason, S.~Dawson, and R.K.~Ellis, Nucl. Phys. {\bf B~327}, 49 (1988);
W.~Beenakker {\it et al.}, Nucl. Phys. {\bf B~351}, 507 (1991);
E.~Laenen, J.~Smith, and W.L. van Neerven, Nucl. Phys. {\bf B~369}, 543
(1992) and Phys. Lett. {\bf B~321}, 254 (1994);
M.~Mangano, P.~Nason, and G.~Ridolfi, Nucl. Phys. {\bf B~373}, 295 (1992);
S.~Frixione {\it et al.}, Phys. Lett. {\bf B~351}, 555 (1995);
N.~Kidonakis and J.~Smith, Phys. Rev. {\bf D~51}, 6092 (1995);
E.~Reya and P.~Zerwas, {\it Proceedings of the ECFA
Workshop on LHC Physics}, Aachen, FRG, 1990, Vol.~II, p.~296.
%
\bibitem{ZGAM}
U.~Baur and E.~L.~Berger, Phys. Rev. {\bf D~47}, 4889 (1993).
%
\bibitem{GREG}
G.~Landsberg, Proceedings of the {\it Workshop on Physics at Current
Accelerators and Supercolliders}, Argonne, Illinois, 1993, edited by
J.~L.~Hewett, A.~R.~White, and D.~Zeppenfeld, p.~303.
%
\bibitem{Sek}
R.~L.~Sekulin, Phys. Lett. {\bf B~338}, 369 (1994).
%
\bibitem{NLCWW}
P.~M\"attig {\it et al.}, Proceedings of the Workshop {\it
$e^+e^-$ Collisions at 500~GeV: The Physics Potential}, Munich,
Annecy, Hamburg, 1991, Vol.~A, p.~223; M.~Bilenky, {\it et al.},
Nucl. Phys. {\bf B~419}, 240 (1994); T.~Barklow, Proceedings
of the Workshop {\it Physics and Experiments with Linear Colliders},
Saariselka, Finland, 1991, Vol.~I, p.~423;
T.~Barklow, SLAC-PUB-6618 (preprint, August 1994), to appear in the
Proceedings of the {\it ``DPF'94 Conference''}, Albuquerque, NM, August
1994.
%
\bibitem{LEPII}
S.~Myers, CERN-SL-95-066 (preprint, June~1995) and talk given at the 2nd
General Meeting of the LEP~II Workshop, CERN, June~15 --~16, 1995.
%
\end{references}
\end{document}